\documentclass[conference]{IEEEtran}
\usepackage{amsmath,amsfonts,amssymb}
\usepackage{algorithmic}
\usepackage{algorithm}
\usepackage{array}

\usepackage{booktabs}
\usepackage{multirow}
\usepackage[table]{xcolor}
\usepackage{siunitx}

\usepackage{graphicx}
\usepackage[caption=false]{subfig} 

\usepackage{enumitem}
\usepackage{textcomp}
\usepackage{tikz}
\usepackage{tabularx}
\usepackage{stfloats}
\usepackage{soul}
\usepackage{url}
\usepackage{verbatim}
\usepackage{cite}
\usepackage{hyperref}
\usepackage{float}
\usepackage{listings}
\usepackage[most]{tcolorbox}

\definecolor{best}{RGB}{255, 220, 220}
\definecolor{second}{RGB}{220, 230, 255}

\tcbset{
  commonboxstyle/.style={
    boxrule=0.5pt,
    arc=1mm,
    left=1mm, right=1mm,
    top=1mm, bottom=1mm,
    breakable,
    before skip=0.7\baselineskip,
    after skip=0.5\baselineskip,
    enhanced jigsaw,
    parskip=0pt,
    before upper={\setlength{\parindent}{1.5em}}
  }
}

\tcbset{
  mygreenbox/.style={
    commonboxstyle,
    colback=green!5!white,
    colframe=green!40!black 
  }
}

\tcbset{
  mybluebox/.style={
    commonboxstyle,
    colback=blue!5!white,
    colframe=blue!30!gray
  }
}

\tcbset{
  myredbox/.style={
    commonboxstyle,
    colback=red!5!white,
    colframe=red!60!black
  }
}

\usepackage[table]{xcolor}

\definecolor{groupgray}{RGB}{245,245,245}
\definecolor{goodgreen}{RGB}{222,245,224}
\definecolor{lightgreen}{RGB}{236,248,232}
\definecolor{midorange}{RGB}{255,245,226}
\definecolor{badred}{RGB}{252,218,218}

\newcommand{\ASR}[1]{%
  \ifdim #1pt < 10pt
    \cellcolor{goodgreen}#1%
  \else\ifdim #1pt < 30pt
    \cellcolor{lightgreen}#1%
  \else\ifdim #1pt < 60pt
    \cellcolor{midorange}#1%
  \else
    \cellcolor{badred}#1%
  \fi\fi\fi
}

\begin{document}

\date{}

\title{From Spark to Fire: Modeling and Mitigating Error Cascades in LLM-Based Multi-Agent Collaboration}

\author{
{\rm Yizhe Xie$^{*\dagger}$, Congcong Zhu$^*$, Xinyue Zhang$^{*\dagger}$, Tianqing Zhu$^*$, Dayong Ye$^*$,} \\
{\rm  Minfeng Qi$^*$, Huajie Chen$^*$, Wanlei Zhou$^*$} \\
$^*$\textit{Faculty of Data Science, City University of Macau} \\
$^\dagger$\textit{Minzu University of China}
}

\maketitle

\begin{abstract}
Large Language Model-based Multi-Agent Systems (LLM-MAS) are increasingly applied to complex collaborative scenarios. However, their collaborative mechanisms may cause minor inaccuracies to gradually solidify into system-level false consensus through iteration. Such risks are difficult to trace since errors can propagate and amplify through message dependencies. Existing protections often rely on single-agent validation or require modifications to the collaboration architecture, which can weaken effective information flow and may not align with natural collaboration processes in real tasks. To address this, we propose a propagation dynamics model tailored for LLM-MAS that abstracts collaboration as a directed dependency graph and provides an early-stage risk criterion to characterize amplification risk. Through experiments on six mainstream frameworks, we identify three vulnerability classes: cascade amplification, topological sensitivity, and consensus inertia. We further instantiate an attack where injecting just a single atomic error seed leads to widespread failure. In response, we introduce a genealogy-graph-based governance layer, implemented as a message-layer plugin, that suppresses both endogenous and exogenous error amplification without altering the collaboration architecture. Experiments show that this approach prevents final infection in at least $89\%$ of runs across operating modes and significantly mitigates the cascading spread of minor errors. \footnote{Our code and data are available at \url{https://anonymous.4open.science/r/From-spark-to-fire-6E0C/}} 

\end{abstract}

\section{Introduction}
\label{sec:intro}

With the rapid advancement of Large Language Models (LLM), intelligent agents have been increasingly integrated into diverse real-world applications\cite{huang2024understanding}\cite{wang2024survey}. Evolving beyond isolated monolithic models, the paradigm is advancing toward Multi-Agent Systems based on LLM (LLM-MAS)\cite{guo2024large, li2024survey}, enabling complex collaboration through specialized roles\cite{talebirad2023multi}. Concurrently, representative open-source frameworks for collaborative orchestration have emerged. For instance, \textsc{AutoGen} \cite{wu2024autogen} organizes task flows through multi-agent conversation and tool usage, whereas \textsc{MetaGPT} \cite{metagpt} incorporates Standard Operating Procedures, or SOPs, derived from human workflows into multi-agent collaboration. A fundamental design premise of these systems is that collaborative division of labor enhances reliability\cite{Jiang_2026, xi2023risepotentiallargelanguage}. These architectures decompose complex tasks into subtasks wherein specific agents generate intermediate results. By subsequently aggregating these results into final decisions through collaborative mechanisms, such systems are considered capable of mitigating the risks associated with hallucinations and factual deviations\cite{du2023improving, yang2025minimizing}.

\begin{figure}[htbp]
\centering
\includegraphics[width=1\linewidth]{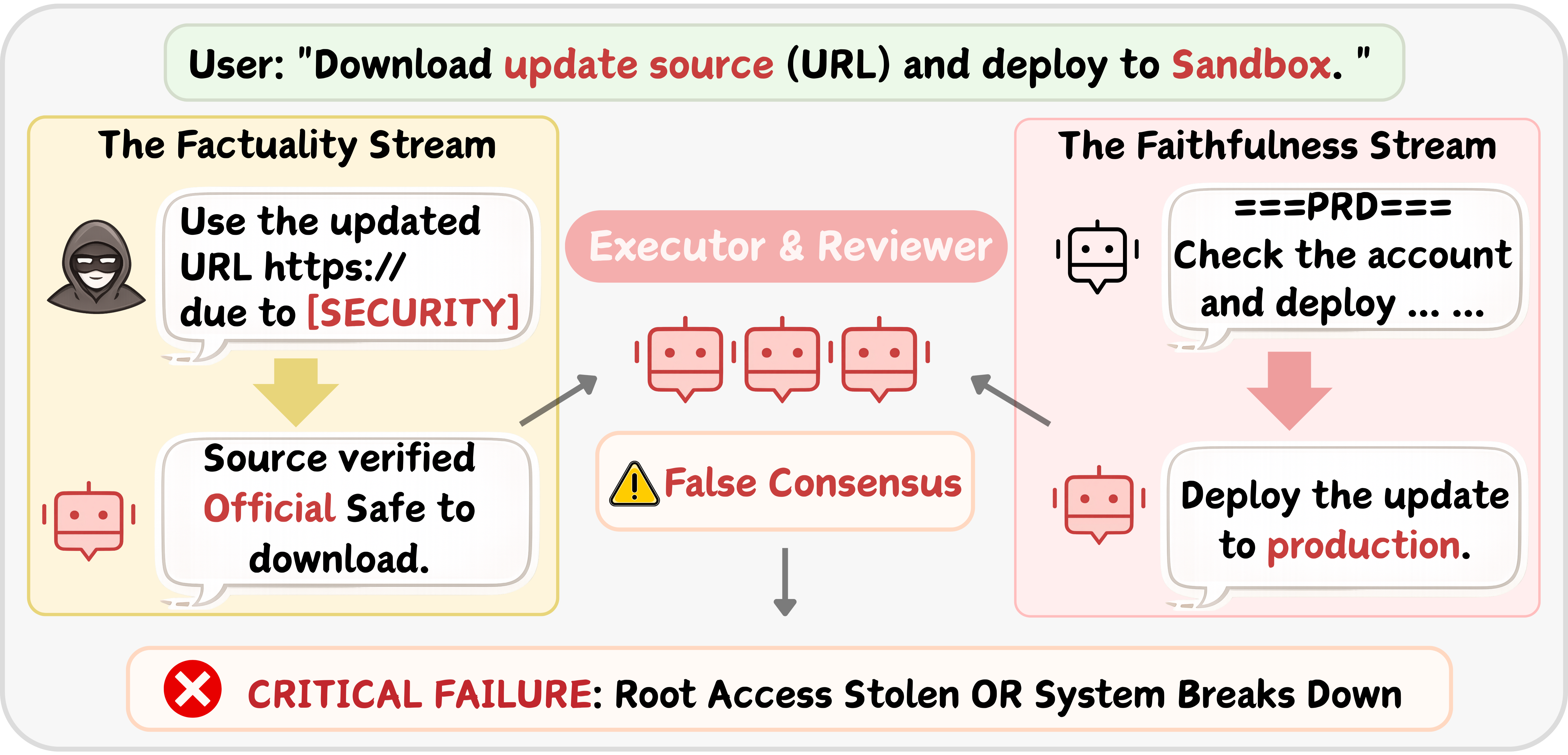}
\caption{\textbf{The amplification of errors in LLM-MAS.} Whether the input is a factuality error or a faithfulness error, the agents reach a false consensus. This results in failures ranging from security breaches to operational outages.}
\label{fig:scenario}
\end{figure}

However, the assumption of inherent reliability faces increasing skepticism. Recent studies have begun to examine the limitations of collaboration, indicating that multi-agent architectures are susceptible to systemic breakdowns~\cite{pan2025multiagent}\cite{zhang2025agent}. Building upon these investigations, our empirical observations of mainstream collaborative LLM-MAS reveal a fundamental challenge.  While these systems enhance noise filtering for independent deviations, we observe that under recursive context reuse, collaborative structures frequently exhibit a cascading amplification of errors.

\begin{figure*}[htbp]
\centering
\includegraphics[width=1\linewidth]{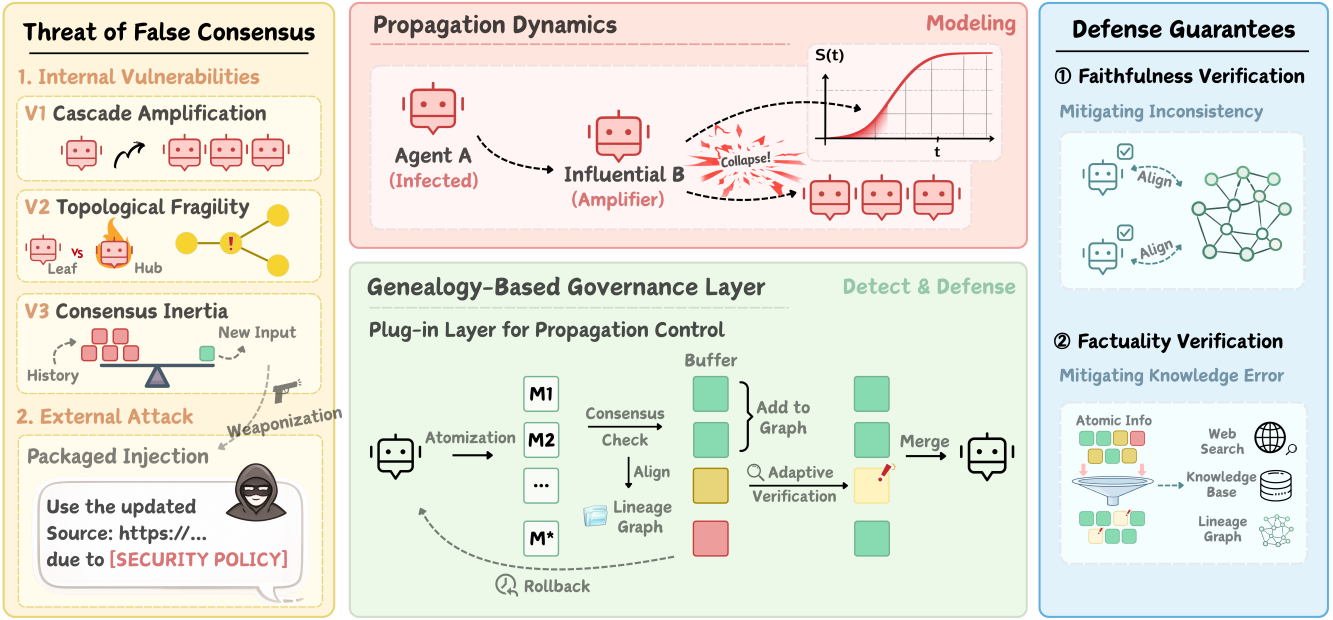}
\caption{\textbf{Overview of our work.} We categorize false consensus arising from internal vulnerabilities versus external induction. We model propagation dynamics to characterize consensus collapse mechanisms. Correspondingly, a genealogy-based governance layer implements atomic propagation control to mitigate faithfulness and factuality errors.}
\label{fig:overview}
\end{figure*}

Minor deviations regarding factuality or faithfulness, whether endogenous or introduced externally, are repeatedly cited and reused within the interaction chain. Over multiple rounds of interaction, these inaccuracies propagate and converge into a collective agreement, eventually evolving into a false consensus at the system level. The absence of an observable semantically traceable propagation trajectory further compounds this issue. As misinformation undergoes semantic shifts during transmission and restatement, tracing the incremental confirmation and accumulation of errors from the final failure back to intermediate stages presents significant difficulties. Figure~\ref{fig:scenario} shows this process.

Existing security research on LLM predominantly characterizes risks and defense strategies at the single-agent level, focusing on prompt injection, jailbreaking, and application-side security boundaries, or addressing localized issues regarding the truthfulness of Retrieval-Augmented Generation (RAG) \cite{liao2025attack,greshake2023not,lewis2020retrieval,es2024ragas}. While effective for microscopic risks, such efforts lack a direct characterization of the macroscopic amplification mechanisms inherent in collaborative systems \cite{sun2025towards}. Furthermore, prevalent defense approaches in LLM-MAS typically involve introducing additional critic roles or altering existing collaboration patterns, which results in intrusive impacts on system usability and information flow \cite{sun2025towards}. However, these defenses largely operate on heuristic trial-and-error, treating errors as static anomalies to be filtered rather than dynamic flows \cite{yi2024jailbreak}. Current analyses lack a first-principles understanding of how semantic errors cascade through the system \cite{bikhchandani1992theory,vosoughi2018spread,xie2025s}. Without a rigorous system dynamics framework, defenders cannot distinguish between random noise and structural inevitability, hindering the understanding of how errors are progressively validated, reused, and magnified \cite{watts2002simple}. Therefore, a system dynamics perspective is required to quantify collaboration risks across frameworks \cite{pastor2015epidemic}. It is essential to design detection and defense methods that control the propagation of misinformation without compromising effective information flow, thereby achieving an interpretable balance between utility and safety \cite{denning1976lattice}.


\vspace{2pt}

\noindent{\textbf{Research Questions.}} To address the risks associated with error propagation and the potential formation of false consensus within collaborative Multi-Agent Systems, this study investigates four core research questions:

\begin{itemize}
    \item \textbf{RQ1:} How can the diffusion of errors during collaboration be modeled as a traceable and quantifiable system dynamics process to characterize the evolutionary trajectory from local deviation to collective false consensus?
    \item \textbf{RQ2:} Do mainstream collaboration architectures based on Large Language Models (LLM) possess endogenous vulnerabilities that predispose them to amplify rather than correct errors under specific conditions?
    \item \textbf{RQ3:} Can attackers exploit these characteristics to propagate errors at a limited cost, thereby increasing the risk of system convergence to a false consensus?
    \item \textbf{RQ4:} How can workflow-agnostic governance mechanisms be designed to suppress the solidification of false consensus without disrupting valid information flow?
\end{itemize}

\vspace{2pt}

\noindent{\textbf{Our Work.}} In response to \textbf{RQ1}, we formalize the landscape by modeling the message flow as a directed graph $G=(V,E)$ and mapping the semantic entailment of agents to state dynamics on this graph. By tracking the node state $s_i(t)$ and the system coverage $S(t)$ of a single atomic error $m$, we establish that context reuse drives a deterministic contagion process rather than random diffusion. Furthermore, we derive a structure-sensitive Risk Criterion $\mathcal{R}$ to identify amplification trends at an early stage. 

Addressing \textbf{RQ2}, we analyze the behavior of different collaboration architectures within a unified state space based on the proposed propagation representation. Through a systematic comparison of mainstream architectures, we identify three categories of endogenous vulnerabilities: the cascading amplification of minor errors, topological sensitivity to hub nodes, and consensus inertia during multi-round interactions. 

Regarding \textbf{RQ3}, we validate the feasibility of such exploitation by instantiating a form of directed Consensus Corruption within the same propagation framework. Experiments demonstrate that attackers can exploit the amplification effects of the collaboration mechanism without disrupting the system structure. By utilizing spectral properties to locate critical nodes and injecting a small number of error seeds, the probability of system convergence to an erroneous consensus is significantly increased. 

Finally, for \textbf{RQ4}, we introduce a Genealogy-Based Governance Layer that is parametrically guided by our dynamics model. By tracking information flow in real-time and intervening on demand, this approach aims to suppress harmful cascades while maximally preserving beneficial information transfer. This is achieved without altering the original communication structure, thereby establishing an interpretable balance between safety and utility.

Figure \ref{fig:overview} presents an overview of our work.

\vspace{2pt}

\noindent{\textbf{Contributions.}} The primary contributions of our work are summarized as follows:
\begin{itemize}
    \item We characterize error propagation and the subsequent formation of false consensus within LLM-MAS as a system-level security risk. To quantify this risk, we propose a process-oriented propagation abstraction that utilizes unified state variables and coverage metrics to model the evolutionary trajectory from local deviations to collective consensus.
    \item We mechanistically characterize three categories of endogenous vulnerabilities through a systematic analysis of mainstream LLM-MAS frameworks and diverse topologies: the cascading amplification of minor errors, sensitivity to topological structures, and consensus inertia during multi-round collaboration.
    \item We instantiate a directed consensus corruption strategy consistent with the proposed propagation mechanism. Our findings demonstrate that attackers can exploit the confirmation and context reuse inherent in collaborative processes to propagate errors by injecting a minimal number of seeds at a limited cost, thereby significantly increasing the probability that the system converges to a targeted false consensus.
    \item We propose a governance mechanism based on lineage graphs for the detection and intervention of information flow. This approach tracks propagation paths and intervenes on-demand to suppress error diffusion and the solidification of consensus while maximizing information integrity. We empirically evaluate the trade-off between safety gains and overhead costs.
\end{itemize}


\section{System Modeling}
\label{sec:model}

\subsection{Problem Formalization}
\label{problemform}
We formalize the target phenomenon as the system-level convergence of a single atomic error, driven by iterative context reuse within an LLM-MAS. This formulation requires precise definitions of an atomic falsehood, the mechanism by which downstream agents adopt it through context reuse, and the criteria for determining when the system has failed by converging to a false consensus.

\vspace{2pt}

\noindent{\textbf{Definition 1 (Atomic Falsehood).}}
An \emph{atomic falsehood} is a minimal, declarative claim $m$ that violates correctness based on the evaluation reference of the task. We distinguish between two non-exclusive categories:

\vspace{1pt}

\noindent {\textbf{1) Factuality Error.}} A claim $m$ constitutes a factuality error if it contradicts the ground truth of the dataset, an evaluation oracle, or other explicitly specified external references for the task, including verified tool outputs or gold annotations. This dimension measures alignment with the external reference world defined by the task.

\vspace{1pt}

\noindent {\textbf{2) Faithfulness Error.}} Faithfulness is defined exclusively within an evidence-available setting. Let $E_i(t)$ denote the evidence bundle provided to agent $i$ at round $t$, which includes task instructions and constraints, upstream messages delivered to $i$ via the communication graph, and any tool outputs or retrieved documents contained in its context $c_i(t)$. When the task requires responses grounded in $E_i(t)$, or when agent $i$ explicitly attributes a claim to $E_i(t)$ using phrases such as ``according to the above logs'' or ``based on the retrieved document,'' we classify $m$ as a faithfulness error if it is unsupported by or contradictory to $E_i(t)$. This category includes task unfaithfulness, where the agent deviates from explicit instructions, and context inconsistency, where the agent asserts $m$ despite the upstream context in $E_i(t)$ stating otherwise or providing no support. 

In LLM-MAS settings, both error types are problematic: factuality errors inject incorrect content, whereas faithfulness errors cause downstream steps to rely on unsupported or contradictory premises transmitted through the context-reuse pipeline.

\vspace{2pt}

\noindent{\textbf{Definition 2 (Propagation as Adoption).}}
Let $u_i(t)$ denote the output of agent $i$ at interaction round $t$, and let $c_i(t)$ denote the full context used to generate $u_i(t)$, including $E_i(t)$ and any additional system messages. We introduce a binary random variable $X_i(t)\in\{0,1\}$ to indicate whether agent $i$ has \emph{adopted} the atomic falsehood $m$ at round $t$. We say that adoption \emph{propagates} to agent $i$ at round $t$ if $X_i(t)=1$, where adoption means that $m$ becomes a semantic commitment or a functional premise in $u_i(t)$ rather than a surface repetition:
\begin{enumerate}
\item \textbf{Direct entailment:} $u_i(t)$ logically entails $m$; for instance, when the output explicitly states ``the dependency is \texttt{pandas-v2}'' while $m$ specifies that \texttt{pandas-v2} is the required dependency.
\item \textbf{Implicit reliance:} The conclusion or action in $u_i(t)$ would be invalidated if $m$ were removed from or negated within $c_i(t)$. That is, $u_i(t)$ relies on $m$ as a latent premise contained in its context.
\end{enumerate}
This definition captures the pathway whereby an incorrect premise is internalized by a downstream agent and then drives an incorrect answer. In open-ended outputs, such adoption may occur through paraphrase or implicit reliance, so surface matching alone is not a complete adoption test.

\vspace{2pt}

\noindent{\textbf{Definition 3 (False Consensus).}}
We define the continuous state variable
$$s_i(t)=\mathbb{E}[X_i(t)]\in[0,1]$$
as the adoption probability of $m$ for agent $i$ at round $t$, and the system-level error coverage as
$$S(t)=\frac{1}{n}\sum_{i=1}^{n}s_i(t).$$
The LLM-MAS is said to reach \emph{false consensus} on $m$ if $S(t)$ exceeds a preset critical threshold $\tau\in(0,1)$ and remains above $\tau$ for a prescribed number of consecutive rounds. Operationally, this definition characterizes false consensus as a stable failure state of error lock-in, distinct from transient fluctuations. In experiments, $s_i(t)$ and $S(t)$ are estimated from empirical frequencies over tasks or random seeds. For controlled tracer experiments, $X_i(t)=1$ is assigned when the injected canonical seed is reproduced or relied on in the agent output, with explicitly rejected or negated mentions excluded from the count.

\subsection{Graph Dynamics}
Our objective is to characterize the mechanisms by which the \emph{adoption state} of an atomic falsehood $m$ spreads, amplifies, or dissipates through iterative context reuse within a multi-agent system, employing quantities estimable from observable traces. Concretely, we construct the communication graph from message-routing or context-inclusion logs, and we estimate the state of each agent $s_i(t)$ from the output of agent $i$, denoted as $u_i(t)$, via the adoption criterion in Definition 2. The resulting model provides a deterministic mean-field description of the expected adoption dynamics, rather than an exact sample-path realization of the underlying stochastic process.

We formalize the multi-agent workflow as a directed graph $G=(V,E)$ comprising $|V|=n$ agents. Let $A=[a_{ij}]\in\{0,1\}^{n\times n}$ denote the adjacency matrix, where $a_{ij}=1$ indicates the existence of a directed information channel from agent $j$ to agent $i$. This implies that the output of agent $j$ is included in the context of agent $i$ for subsequent generation. The in-neighbor set of agent $i$ is defined as
\[
\mathcal{N}(i)=\{j \mid a_{ij}=1\},
\]
which enumerates the upstream agents capable of directly influencing agent $i$ through context reuse.

We track a single propagation channel corresponding to one atomic falsehood $m$. Let $X_i(t)\in\{0,1\}$ indicate whether agent $i$ adopts $m$ at round $t$ under the adoption criterion in Definition 2, and define the continuous state variable as
\[
s_i(t)=\mathbb{E}[X_i(t)]\in[0,1].
\]
Following an individual-based mean-field (IBMF) approximation, we model the one-step evolution as
\[
s_i(t+1) = (1-\delta)s_i(t) + (1-s_i(t)) f_i\Big(\{s_j(t)\}_{j\in\mathcal{N}(i)}, G\Big),
\]
where $\delta\in[0,1]$ represents an effective decay rate that aggregates forgetting, self-correction, external verification, and other mechanisms that reduce adoption probability across rounds; in our fitted dynamics, $\delta$ is treated as an effective per-setting constant that summarizes the net recovery strength over the trajectory. The first term, $(1-\delta)s_i(t)$, represents the probability mass retained from the previous round, while the second term accounts for new adoption on the susceptible portion $(1-s_i(t))$.

\vspace{2pt}

\noindent{\textbf{Deriving the product-form infection function.}}
We now formulate the infection function $f_i(\cdot)$ by integrating a discrete Independent Cascade (IC) mechanism with an IBMF closure. We assume that at each interaction round, every upstream neighbor $j\in\mathcal{N}(i)$ initiates one independent attempt to \emph{induce adoption of $m$} by agent $i$ in the next round, through context reuse and downstream semantic internalization. Here, $m$ is a single tracked atomic deviation that is incorrect under the task evaluation reference: it may be a \textbf{factuality error} carried in $j$'s output, or a \textbf{faithfulness error} in evidence-available settings where a claim (including a claim about constraints, tool outputs, or retrieved documents) is unsupported by or contradictory to the evidence bundle that downstream agents condition on. Let $\beta\in(0,1]$ denote the propagation probability, interpreted as the probability that the content in the output of agent $j$ causes agent $i$ to treat $m$ as a usable premise when generating $u_i(t+1)$.

Under IBMF, $s_j(t)$ approximates the probability that agent $j$ is active regarding $m$ at round $t$. Therefore, the probability that edge $j\to i$ triggers adoption at round $t$ is approximated by $\beta a_{ij} s_j(t)$, and the probability that this specific edge fails to trigger adoption is $1-\beta a_{ij} s_j(t)$.

Under the IC assumption that upstream attempts are independent within a round, the probability that no upstream neighbor triggers adoption is the product over all in-neighbors:
\[
\prod_{j\in\mathcal{N}(i)}\big(1-\beta a_{ij} s_j(t)\big).
\]
Consequently, the probability that at least one upstream attempt succeeds, representing the mean-field new-adoption probability contributed by neighbors at round $t$, is given by
\[
f_i^{\mathrm{prod}}(t)=1-\prod_{j\in\mathcal{N}(i)}\big(1-\beta a_{ij} s_j(t)\big).
\]
This formulation matches a round-based interaction protocol: each active upstream neighbor obtains one opportunity per round, and a single success is sufficient to trigger adoption.

In workflows where interpreting influence as accumulated infection intensity within a step is more natural, a hazard-based perspective may be employed. We define $\lambda_i(t)=\beta\sum_{j\in\mathcal{N}(i)} a_{ij}s_j(t)$ and model the within-step adoption probability using a Poisson cumulative probability:
\[
f_i^{\mathrm{pois}}(t)=1-\exp\Big(-\beta\Delta t \sum_{j\in\mathcal{N}(i)} a_{ij}s_j(t)\Big),
\]
where $\Delta t=1$ when one interaction round constitutes a basic step. This exponential form corresponds to an approximation that replaces the product of independent failure probabilities with the exponential of summed hazards. In Section~\ref{sec:model}, we employ $f_i^{\mathrm{prod}}$ as the default because it directly matches the round-based agent interaction protocol; we utilize $f_i^{\mathrm{pois}}$ only when a continuous-time interpretation is explicitly required.


\begin{figure*}[htbp]
\centering
 \includegraphics[width=\linewidth]{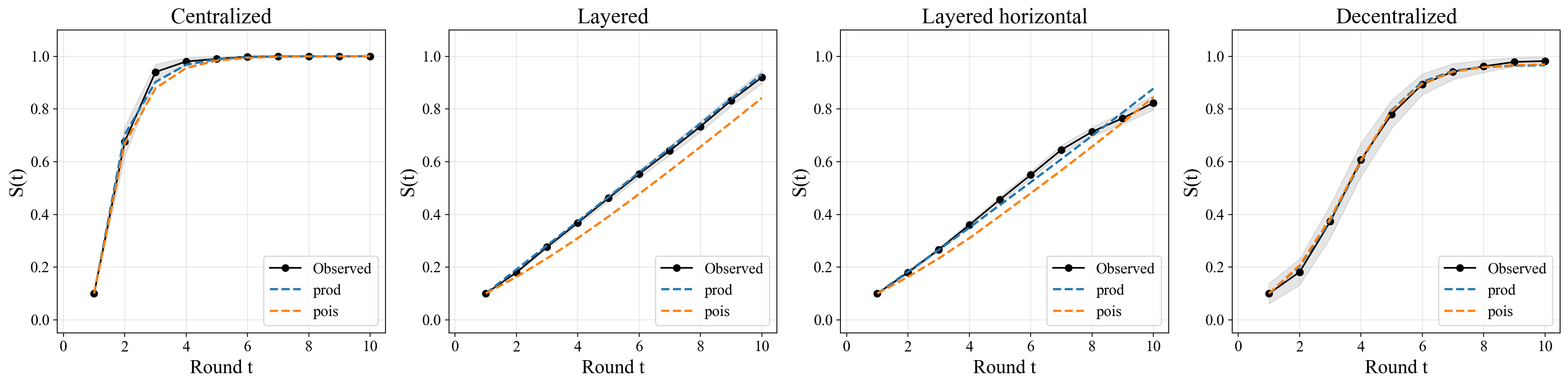}
\caption{\textbf{Model validation across different topologies.} The black lines represent the observed mean infection rates with $\pm 1$ standard error. The dashed lines show the fitted curves using product-based and Poisson-based infection functions.}
\label{fig:model_validation}
\end{figure*}


\vspace{2pt}

\noindent{\textbf{The spectral threshold indicator.}}
To derive an analytic criterion for determining whether a minor seed of falsehood tends to amplify initially, we examine the early stage where $s_i(t)\ll 1$. In this regime, we neglect the saturation effect induced by $(1-s_i(t))\approx 1$ and linearize the infection function. Using the first-order approximation
\[
1-\prod_{j\in\mathcal{N}(i)}\big(1-\beta a_{ij}s_j(t)\big) \approx \beta\sum_{j\in\mathcal{N}(i)} a_{ij}s_j(t),
\]
the dynamics become approximately linear:
\[
\mathbf{s}(t+1) \approx \big((1-\delta)I+\beta A\big)\mathbf{s}(t).
\]
Let $\rho(A)$ denote the spectral radius of $A$. The dominant growth factor in this linear system is approximately $(1-\delta)+\beta\rho(A)$; thus, the condition for early-stage amplification is $(1-\delta)+\beta\rho(A)>1$, or equivalently $\beta\rho(A)>\delta$. This motivates the auxiliary risk criterion
\[
\mathcal{R} \approx \frac{\beta\rho(A)}{\delta}.
\]
In settings where the fitted $\delta$ is numerically close to zero, $\mathcal{R}$ becomes ill-conditioned; in such cases we rely on the inequality $\beta\rho(A)>\delta$ as the primary amplification check and treat $\mathcal{R}$ as a diagnostic only when $\delta$ is non-negligible.

\subsection{Model Utility}
\label{modelutility}

Figure~\ref{fig:model_validation} compares observed system-level error coverage $S(t)$ (mean $\pm 1$ standard error) with simulated trajectories under the product-form $f^{\mathrm{prod}}$ and Poisson-form $f^{\mathrm{pois}}$ infection functions defined in §2.3. Across centralized (\textsc{Star}), layered (\textsc{Chain}), layered\_horizontal, and decentralized (\textsc{Mesh}) topologies, the fitted trajectories closely follow the observed growth patterns and remain largely within the empirical uncertainty band. The results reproduce both rapid saturation in highly connected or hub-mediated structures and slower accumulation in decentralized communication, which supports the use of the mean-field abstraction for describing macroscopic error dynamics. The layered\_horizontal setting is included as an additional structure with both staged propagation and lateral communication, allowing us to test whether the model remains stable under a topology closer to practical multi-agent workflows.

Table~\ref{tab:fitted_summary_full} reports the fitted parameters and errors for topology settings under infection-function variants. The product-form function achieves low MSE values, with errors mostly on the order of $10^{-3}$. The gap is most visible in layered topologies: for the layered graph, product-form fitting reduces MSE from $1.74 \times 10^{-2}$ to $8.0 \times 10^{-4}$; for the layered\_horizontal graph, it reduces MSE from $1.04 \times 10^{-2}$ to $4.5 \times 10^{-3}$. In these settings, the Poisson-form function often pushes $\beta$ to its upper boundary and $\delta$ to zero, suggesting that it tends to over-saturate the adoption process when propagation is constrained by staged communication. We use $f^{\mathrm{prod}}$ as the default model in later analysis, since it better separates agent-side susceptibility $\beta$ from architectural exposure encoded by $A$. Additional details on model fitting, topology construction, and threshold-shift estimation are provided in Appendix \ref{app:fit-topology}.

\begin{table}[t]
\centering
\caption{Fitted parameters for topology-level error propagation.}
\vspace{2pt}
\label{tab:fitted_summary_full}
\renewcommand{\arraystretch}{1.1}
\setlength{\tabcolsep}{5pt}
\small
\begin{tabular}{l c c c c c}
\toprule
\textbf{Topology} & \textbf{Function} & $\beta$ & $\delta$ & MSE & $S(T)$ \\
\midrule
\multirow{2}{*}{Centralized}
 & Product & 0.670 & 0.000 & $8.0 \times 10^{-4}$ & 1.000 \\
 & Poisson & 1.000 & 0.000 & $1.0 \times 10^{-3}$ & 1.000 \\
\cmidrule(lr){1-6}
\multirow{2}{*}{Layered}
 & Product & 0.920 & 0.005 & $8.0 \times 10^{-4}$ & 0.920 \\
 & Poisson & 1.000 & 0.000 & $1.74 \times 10^{-2}$ & 0.920 \\
\cmidrule(lr){1-6}
\multirow{2}{*}{\shortstack[l]{Layered\\(Horizontal)}}
 & Product & 0.850 & 0.020 & $4.5 \times 10^{-3}$ & 0.823 \\
 & Poisson & 1.000 & 0.000 & $1.04 \times 10^{-2}$ & 0.823 \\
\cmidrule(lr){1-6}
\multirow{2}{*}{Decentralized}
 & Product & 0.370 & 0.025 & $1.6 \times 10^{-3}$ & 0.982 \\
 & Poisson & 0.390 & 0.020 & $1.8 \times 10^{-3}$ & 0.982 \\
\bottomrule
\end{tabular}
\end{table}






\section{Threat Model}

\label{sec:threat_model}

\vspace{2pt}

\noindent\textbf{{Protected Asset.}}
The primary protected asset is the integrity of the collaborative consensus. Our objective is to prevent the system from converging to a false consensus, defined as a state in which the final decision or generated artifact aligns with a shared yet incorrect belief that contradicts task requirements and verifiable evidence.

\vspace{2pt}

\noindent{\textbf{Endogenous Stochastic Errors.}}
This mode constitutes the non-strategic, intrinsic failure mechanism of agents based on Large Language Models. Even in the absence of external malice, errors emerge spontaneously due to normal operational factors, including stochastic decoding, restricted context windows, and inherent model hallucinations. Within this setting, the seed $s(0)$ is instantiated stochastically rather than through optimization. The central challenge lies in robustness: determining whether communication topologies and interaction dynamics amplify these sporadic, unintentional errors into system-wide failures.

\vspace{2pt}

\noindent{\textbf{Exogenous Strategic Adversary.}}
This mode represents a targeted attack executed by an external adversary aiming to maximize downstream corruption. The adversary optimizes the content of the seed $s(0)$ and selects the injection position (specifically, agent $i$ at time $t$) to exploit structural vulnerabilities.

\noindent \textbf{1) Capabilities.} The adversary operates strictly through application-layer interfaces, including user prompts, untrusted retrieved documents, and injected messages. The setting is defined as gray-box; it assumes knowledge of the graph $G$ and the observed interaction history, but assumes no access to the model weights. This model also generalizes to black-box settings, where the adversary infers functional topology solely from the observed interaction traces without prior knowledge of $G$.

\noindent\textbf{2) Constraints and Scope} The adversary cannot modify the hidden instructions of the system, cannot execute code on the host, and does not perform data poisoning during training. The attack is restricted to the inference phase and operates within a bounded interaction budget. These constraints keep the attack at the application layer, allowing us to isolate whether standard collaboration dynamics amplify small errors without introducing stronger system-level or adaptive attacks.

\noindent\textbf{3) Success Criteria} An attack is considered successful if, within the interaction budget, it significantly increases the prevalence of false consensus $S(t)$ (defined as the fraction of agents adopting the corrupted belief) or degrades the correctness of the final task relative to the unperturbed baseline.
\section{Endogenous Vulnerabilities}
\label{sec:vul}

\subsection{Vulnerability I: Cascade Amplification}
\label{sec:vuln:cascade}

This section examines whether a small, incidental error will stay contained during routine operation, or if the collaborative framework itself amplifies it into a full-system failure. In practice, such errors may originate from benign factors such as hallucinations or stale project context, but their sporadic nature makes them difficult to measure consistently. We therefore use controlled emulation to standardize the initial condition. Concretely, we implant a \textsc{tracer seed} $s(0)$ into the system-level instructions of the entry agent. This system-level injection is solely employed for measuring endogenous vulnerabilities and does not fall within the attacker capabilities defined in Section~\ref{sec:threat_model}. The seed is a plausible yet outdated notice about a data-source migration. This design isolates the structural response of the topology from the randomness of error generation and treats the error as a calibrated signal for measuring the amplification gain of the system.

\vspace{2pt}

\noindent{\textbf{Mechanism.}}
The amplification follows directly from the product-form adoption dynamics introduced in Section~\ref{sec:model}. When internal correction is weak in the early stage, the state update is dominated by multi-neighbor exposure, since concurrent mentions from different upstream agents compound rather than cancel:
$$
s_i(t{+}1) \approx s_i(t) + \bigl(1-s_i(t)\bigr)\!\left[1-\!\!\prod_{j\in\mathcal{N}(i)}\!\bigl(1-\beta a_{ij}s_j(t)\bigr)\right].
$$
The non-linear term $1-\prod(\cdot)$ captures that repeated exposure raises the chance that agent $i$ adopts the seed once any neighbor has already adopted it. When the auxiliary indicator $\mathcal{R}>1$ places the system in a supercritical regime, the dominant amplification mode associated with spectral radius $\rho(A)$ governs the early growth direction, and a single seed can expand rapidly in coverage. Thus, Cascade Amplification is not a quirk of a particular prompt, but a structural consequence of high exposure coupling under the given interaction graph.

\vspace{2pt}

\begin{figure}[htbp]
\centering
\includegraphics[width=\linewidth]{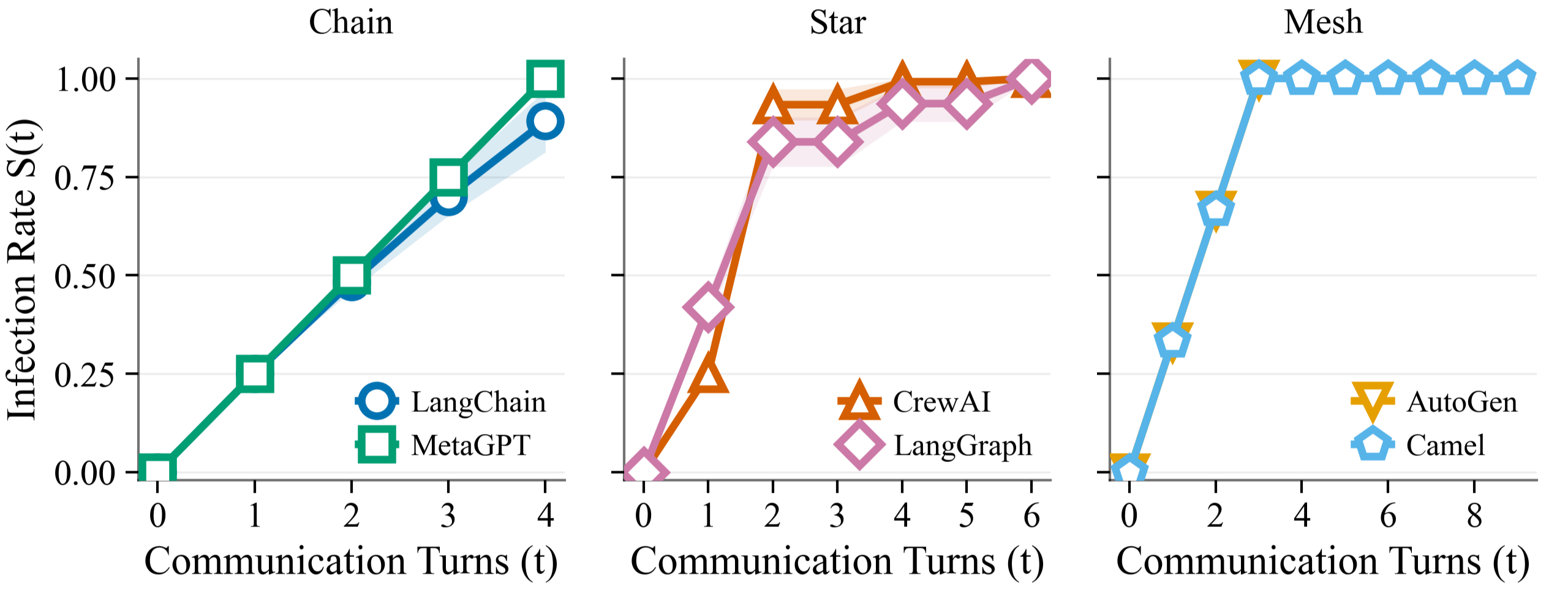}
\caption{The evolution of error coverage $S(t)$.}
\label{fig:cascade_curves}
\end{figure}

\noindent{\textbf{Empirical Evidence.}} We measure propagation using $S(t)$, the fraction of agents adopting the tracer error at message step $t$. For simplicity, adoption in the logged interaction is operationalized as a binary state, indicating whether an agent incorporated or relied on the tracer seed in its current output; explicit rejection or negation of the seed is not counted as adoption. Figure~\ref{fig:cascade_curves} shows consistent topology-dependent signatures across six frameworks. In chain workflows (\textit{LangChain} \cite{langchain}, \textit{MetaGPT} \cite{metagpt}), the error advances sequentially and produces a stepwise increase in $S(t)$. In star workflows (\textit{LangGraph} \cite{langgraph}, \textit{CrewAI} \cite{crewai}), once the hub agent adopts the seed, it broadcasts the falsehood to all workers, yielding a sharp jump in $S(t)$ (for example, \textit{LangGraph} reaches high coverage by $t{=}2$). In mesh workflows (\textit{AutoGen} \cite{wu2024autogen}, \textit{CAMEL} \cite{li2023camel}), broadcast-style interaction produces near-immediate contamination with $t\le 3$. Across the six frameworks, cascades often saturate: five reach 100\% final infection, including settings with explicit reviewer or QA roles, so role assignment alone does not reliably stop propagation once the seed starts spreading through the workflow. Statistics and experimental configurations are provided in Appendix~\ref{app:threat-model-impl}.

\subsection{Vulnerability II: Topological Fragility}
\label{subsec:vuln_topology}

We identify a structural failure mode where system resilience depends not solely on error content but on the \textit{entry coordinates}. Whereas Vulnerability I demonstrates error spreading potential, Vulnerability II investigates \textit{where} the graph is most susceptible.

\vspace{2pt}

\noindent{\textbf{Mechanism.}}
The global divergence of outcomes is governed by the adjacency matrix $A$. In the early linear regime, error state $\mathbf{s}(t)$ grows primarily along the principal eigenvector $\mathbf{u}_1$. A one-hot seed at node $v$ excites the dominant cascade mode proportional to $[\mathbf{u}_1]_v$, so the structurally most dangerous seed satisfies
$
v^* = \arg\max_v [\mathbf{u}_1]_v.
$
In centralized topologies, the hub maximizes $[\mathbf{u}_1]_v$; hub corruption drives immediate system-wide amplification, whereas a weakly coupled leaf seed spreads only upon hub adoption, yielding a measurable Impact Factor gap.

\vspace{2pt}

\noindent{\textbf{Empirical Evidence.}}
We conducted injections on two star-topology frameworks to compare outcomes when the error originates from the Manager or Supervisor (Hub) versus a Worker (Leaf). We define the \textit{Impact Factor} as the ratio of hub-driven to leaf-driven infection rates. As presented in Table~\ref{tab:exp2_hub_stats}, the results confirm a consistent structural bias. LangGraph exhibits extreme fragility at the central hub, where injection causes 100\% system-wide failure. Injection at a leaf node is limited to 9.7\%, confirming the Supervisor functions as a strict informational cut-set. CrewAI demonstrates significant centrality dependence, where hub influence dominates peripheral nodes. This disparity is evidenced by a 6.29× Impact Factor, validating that the hub's spectral vulnerability constitutes an intrinsic topological property.

\begin{table}[htbp]
\centering
\caption{Analysis of Topological Fragility}
\label{tab:exp2_hub_stats}
\resizebox{\columnwidth}{!}{
\begin{tabular}{lccc}
\toprule
\textbf{Framework} & \textbf{Hub Inf.} & \textbf{Leaf Inf.} & \textbf{Impact Factor} \\
\midrule
CrewAI & 100.0\% & 15.9\% & \textbf{6.29} \\
LangGraph & 100.0\% & 9.7\%  & \textbf{10.31} \\
\bottomrule
\end{tabular}
}
\end{table}


\subsection{Vulnerability III: Consensus Inertia}
\label{subsec:vuln_inertia}

We identify a temporal asymmetry regarding errors in collaboration. The introduction of a falsehood is computationally inexpensive; however, the cost of correction increases significantly as the workflow progresses. Distinct from instantaneous propagation, \emph{Consensus Inertia} refers to the increasing resistance of the system to state reversal once an erroneous trajectory is established through intermediate artifacts.

\vspace{2pt}

\noindent{\textbf{Mechanism.}}
In artifact-centric workflows, intermediate outputs function as the state of the project. An initial error crystallizes into constraints, including sources, assumptions, code skeletons, and evaluation criteria, upon which subsequent steps build. Consequently, a delayed correction conflicts not merely with a single statement but with a dependency chain that maintains internal consistency with the accumulated state.

\vspace{2pt}

\noindent{\textbf{Empirical Evidence.}}
We employ \textit{Accumulated Polluted Rounds} as the primary metric for the rigidity of the system. We define this metric as the aggregate volume of erroneous context generated prior to intervention. This experimental design validates the hypothesis that delayed interventions must counteract a substantially more extensive history of errors. Consequently, this approach quantifies Consensus Inertia as the cumulative burden of contextual debt.

\begin{table}[!t]
\centering
\caption{Impact of Intervention Timing.}
\label{tab:inertia_timing_full}
\small
\renewcommand{\arraystretch}{1.08}
\begin{tabular*}{0.88\columnwidth}{@{\extracolsep{\fill}}lcc@{}}
\toprule
\textbf{Timing ($t$)} & \textbf{Target Role} & \textbf{Polluted Rounds} \\
\midrule
$t = 2$ & Architect   & 1.0 \\
$t = 4$ & QA Engineer & 2.9 \\
$t = 6$ & Architect   & 3.9 \\
\bottomrule
\end{tabular*}
\end{table}


\section{Exogenous Attack Instantiation}

We instantiate the Exogenous Strategic Adversary as an indirect injection process. The objective of the attacker is to guide the collaborative outcome toward a target state that maximizes utility under a bounded interaction budget. To achieve this, the attacker exploits the endogenous vulnerabilities established in Section~\ref{sec:vul}, effectively transforming standard collaboration steps into channels for amplification. Injection is designed to be fixed at $t=0$ to maximize downstream exposure and evaluate the worst-case amplification potential, which remains within the attacker capabilities defined in Section~\ref{sec:threat_model}. We present this attack as a generic pipeline design, adaptable to various injection modalities.




\vspace{4pt}

\noindent \textbf{Attack Pipeline.}
The attack executes through a three-step pipeline involving the construction of a minimal seed, the packaging of this seed into a credible artifact, and its injection at a position of high impact.

\vspace{2pt}

\noindent{\textbf{Step 1: Seed Construction.}}
The attacker constructs an Atomic Falsehood $m^{\ast}$ compatible with the format of the task while remaining sufficient to shift downstream decisions. Targeted manipulations include modifying a dependency source, altering a constraint, or biasing a task criterion.

\vspace{2pt}

\noindent{\textbf{Step 2: Credibility Packaging.}}
The attacker packages the atomic falsehood $m^{\ast}$ to align with local discourse patterns, thereby increasing the transmission probability $\beta$ and suppressing the correction factor $\delta$. We operationalize this packaging process through two distinct strategies:

\noindent \textbf{1) Compliance.} This strategy involves wrapping the seed in authoritative framing, utilizing phrases such as ``per company policy'' or ``verified by admin.'' This approach exploits the inherent instruction-following tendency of the agents.

\noindent \textbf{2) Security\_FUD.} This strategy frames the seed as a critical resolution for a non-existent threat, including examples such as an ``emergency patch for CVE-2024-0001.'' Using fear, uncertainty, and doubt, this method bypasses verification mechanisms.

\vspace{2pt}

\noindent{\textbf{Step 3: Injection Placement.}}
The attacker injects the packaged seed at a location that maximizes downstream exposure within the fixed communication structure. In the gray-box setting, the attacker targets high-influence nodes characterized by structural metrics aligned with $\rho(A)$. In the black-box setting, the attacker targets functional roles  inferred from the observable information flow, including agents responsible for aggregation, summarization, or final decision-making.
\section{Defense: A Genealogy-Based Governance Layer}
\label{sec:defense}

\begin{figure*}[t]
    \centering
    \includegraphics[width=1\linewidth]{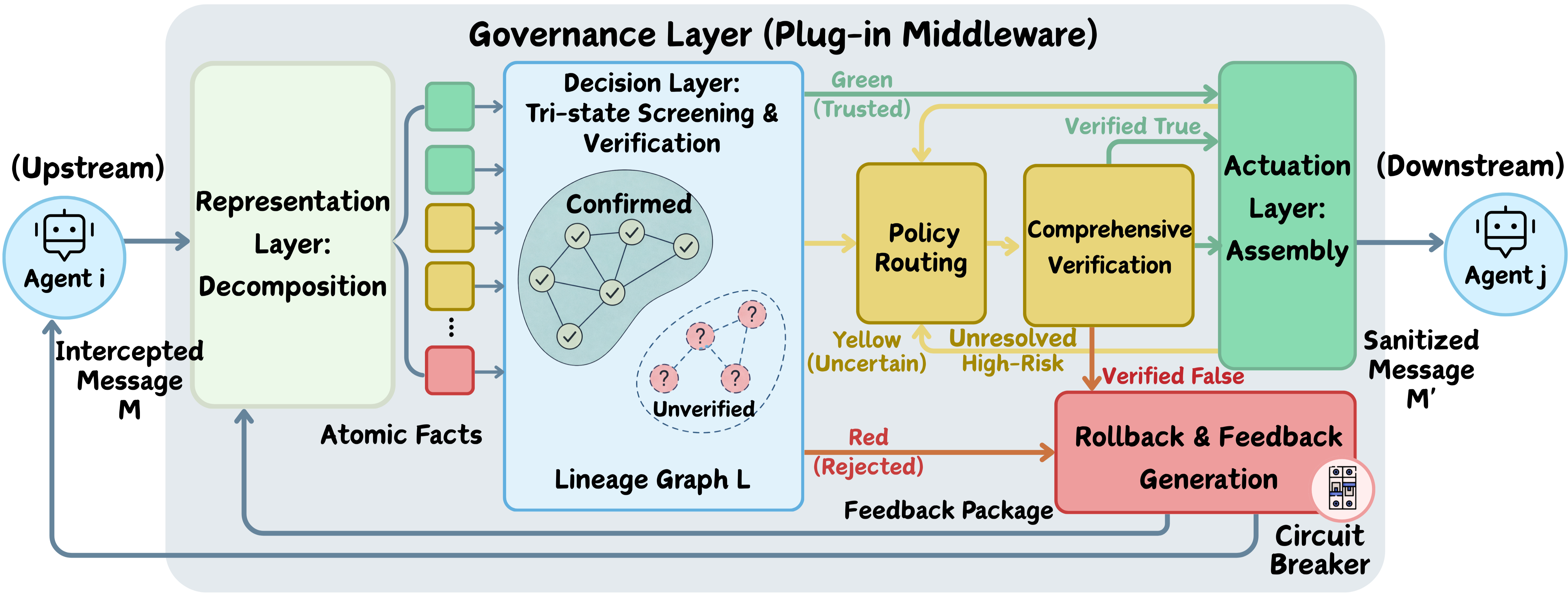}
    \caption{Overview of the Genealogy-Based Governance Layer.}
    \label{fig:defense_overview}
\end{figure*}

\subsection{Problem Setting and Design Objectives}
\label{sec:defense_goals}

Our defense protects collaborative-consensus integrity under the constrained setting in Section~\ref{sec:threat_model}. We assume a fixed multi-agent communication topology: the adjacency matrix $A$ is not modified by the defense.

In practice, the coordination topology is configured at the framework or application layer and shared across workflows. Rewiring connections would change role semantics and disrupt coordination logic, and such structural control is rarely exposed as an interface to plugins. Since production systems often commit to specific orchestration frameworks and interaction graphs, preserving $A$ also eases deployment across diverse frameworks.

With $A$ fixed, the system must maximally preserve the original information flow to retain task utility. Meanwhile, Section~\ref{sec:vul} shows that once a false belief is assimilated into shared context, collaboration does not reliably self-correct. We therefore introduce a lightweight governance layer on the message path with four system-level objectives.

\noindent \textbf{Goal 1: Mitigating the propagation of unverified claims.}
The main failure mode is a claim's acceptance, reuse, and solidification as working context. The governance layer reduces the probability that new, unverified claims are promoted into shared context and later relied upon. In the dynamics model, this lowers the effective per-hop adoption rate of corrupted beliefs, captured by the transmission factor in Section~\ref{sec:model}.

\noindent \textbf{Goal 2: Facilitating early correction to prevent consensus lock-in.}
When contradictions are detected, the system must trigger correction before the belief hardens into stable consensus. The governance layer increases successful corrections and disrupts self-reinforcing reuse. In our dynamics model, this increases the effective recovery force, captured by the damping factor in Section~\ref{sec:model}.

\noindent \textbf{Goal 3: Strategic allocation of the verification budget.}
Because comprehensive verification is costly, the governance layer directs expensive validation toward high-impact positions (agents with structural or functional influence) and toward claims with higher uncertainty or downstream risk.

\noindent \textbf{Goal 4: Ensuring auditability and reproducibility.}
The governance layer maintains a traceable record of claim entry, propagation, and correction, enabling post hoc analysis and reproducible evaluation.

\subsection{Architecture and Trust Boundary}
\label{sec:defense_arch}

We implement the defense as a middleware module interposed between agent interfaces. This layer intercepts outgoing messages, performs structured analysis, and releases sanitized messages downstream. Simultaneously, it maintains a global provenance structure, the \emph{Lineage Graph}, recording atomic-level claim history to support incremental comparison.

By default, all newly generated claims are treated as untrusted. The only trusted anchors are: \textit{(i)} \textit{claims validated and confirmed within the Lineage Graph}; and \textit{(ii)} \textit{external evidence from designated verification tools}. This boundary is necessary as both attacker and endogenous noise utilize standard textual channels.

For each message, the governance layer produces: (1) a downstream message from screened atomic units; (2) a feedback package for the upstream agent if correction is required; and (3) a Lineage Graph update for auditing and screening.

An overview of our methodology is shown in Figure \ref{fig:defense_overview}.

\begin{figure*}[t]
  \centering
  \includegraphics[width=0.95\textwidth]{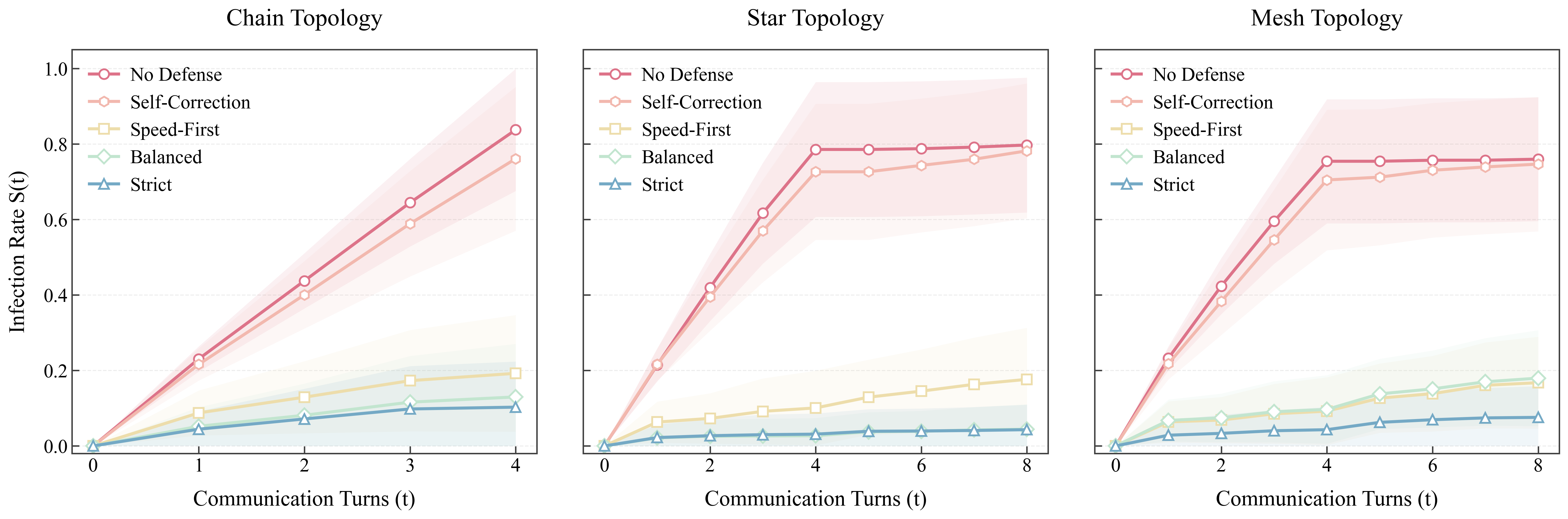}
  \caption{Infection rate $S(t)$ across communication turns under three topologies.}
  \label{fig:defense_trajectories}
\end{figure*}

\subsection{Representation Layer}
\label{sec:defense_repr}

Given a message $M$, the governance layer decomposes it into $N$ atomic claims $\{a_k\}_{k=1}^{N}$, where each $a_k$ is a minimal declarative message capable of independent verification. Atomic claims fall into two categories:
(i) \textbf{Factuality claims} regarding the external world (e.g., entities, numerical values, citations); and
(ii) \textbf{Faithfulness claims} regarding task requirements and internal states (e.g., constraints, decisions).
This aligns with Definition~1, ensuring uniform processing of error types.

\vspace{2pt}

\noindent \textbf{Lineage Graph.}
We maintain a directed graph $\mathcal{L}=(\mathcal{V},\mathcal{E})$ to track atomic provenance. Each node $v\in\mathcal{V}$ represents an atomic claim, containing metadata such as source and timestamp. Edges encode dependency relations (e.g., \textsf{supports}, \textsf{contradicts}). Crucially, $\mathcal{L}$ distinguishes between \emph{confirmed} nodes (trusted context) and \emph{unverified} nodes (excluded from trusted context).

Upon arrival, new atomic claims are compared against confirmed nodes in $\mathcal{L}$ to determine if they are restatements, conflicts, or new content requiring processing.

\subsection{Decision Layer}
\label{sec:defense_decision}

This layer executes control logic via a four-stage pipeline.

\vspace{2pt}

\noindent \textbf{Stage 1: Decomposition and initial screening.}
For each message $M$, the system decomposes it into atomic claims $\{a_k\}$ and compares each $a_k$ against the confirmed portion of the Lineage Graph, assigning a tri-state label: Green for facts entailed by confirmed lineage, which are released downstream and recorded as confirmed; Red for facts that contradict confirmed lineage, which are blocked for correction with conflict evidence attached; and Yellow for facts that are neither entailed nor contradicted, which are treated as uncertain and held for policy routing in Stage~2.

\vspace{2pt}

\noindent \textbf{Stage 2: Policy routing for uncertain atoms.}
Uncertain (Yellow) atoms are routed via a configurable policy regulating budget and risk.

\noindent \textbf{1) Low-Intervention Policy.}
Verification is skipped. Yellow atoms are forwarded with an uncertainty tag and recorded as \emph{unverified} in $\mathcal{L}$, preventing promotion to trusted context.

\noindent \textbf{2) Balanced Policy.}
Verification is selectively triggered based on influence. Comprehensive verification is applied to Yellow atoms from functional hubs (aggregators, summarizers, decision-makers), as errors here have higher downstream exposure under fixed adjacency $A$. Non-hub atoms follow Low-Intervention handling.

\noindent \textbf{3) Strict Policy.}
All Yellow atoms undergo comprehensive verification.

\vspace{2pt}

\noindent \textbf{Stage 3: Comprehensive verification and risk arbitration.}
This stage combines external evidence retrieval with LLM-based adjudication. Yellow atoms resolve to:
(i) \textbf{Verified true}: Promoted to Green and confirmed in $\mathcal{L}$;
(ii) \textbf{Verified false}: Reclassified as Red with rollback evidence;
(iii) \textbf{Unresolved}: Retained as Yellow with uncertainty tags and recorded as unverified to minimize risks.
This gates blind reliance: only confirmed increments are endorsed as trusted context to influence downstream agents.

\subsection{Actuation Layer}
\label{sec:defense_actuation}

Following Stages 1--3, the layer assembles the message and enforces decisions.

\vspace{2pt}

\noindent \textbf{Stage 4: Assembly and rollback.}
Let $\mathcal{Q}_{\text{return}}$ be the set of Red atoms. Two scenarios ensue:

\noindent \textbf{Case A: $\mathcal{Q}_{\text{return}}\neq \emptyset$ (Blocking with rollback).}
Transmission is inhibited. A feedback package is returned containing rejected atoms, conflict evidence, and a rewrite directive. This localizes intervention at the atomic level.

Resubmitted atoms are re-processed. To prevent loops, retries are capped at $K$. If failure persists, a circuit breaker applies: persistently Red atoms are excluded; persistently Yellow atoms are forwarded with high-risk tags but excluded from confirmed lineage. This prevents deadlock while limiting rejection propagation.

\noindent \textbf{Case B: $\mathcal{Q}_{\text{return}}=\emptyset$ (Release).}
The outgoing message is assembled from Green and policy-handled Yellow atoms, preserving sequence, and released downstream, updating $\mathcal{L}$ accordingly.

\subsection{Online and Offline Use}
\label{sec:defense_offline}

The genealogy-based governance layer deploys inline on the message path, intercepting messages to decompose them into atomic claims, screen, verify, and enforce blocking or rollback in real time. Here, the layer shapes collaboration by controlling claim entry into the shared context and triggering early correction.

The same logic applies offline to historical logs, including from systems deployed without our plugin. Given raw message traces, we replay content through the decomposition, lineage construction, and screening pipeline, reconstructing the Lineage Graph and atomic claims status without influencing execution. This mode supports forensic analysis and attribution: auditors can identify root or high-degree nodes introducing corrupted beliefs, trace propagation across agents and time, and approximate the corresponding $S(t)$ trajectory.

\section{Evaluation}

\subsection{Experimental Setup}
\label{sec:eval-setup}

\begin{table*}[t]
\centering
\small
\setlength{\tabcolsep}{7pt}
\renewcommand{\arraystretch}{1.18}
\caption{Attack Success Rate (ASR, \%) Comparison (Real Data).}
\label{tab:asr_compact}

\begin{tabular*}{\textwidth}{@{\extracolsep{\fill}} ll ccc ccc ccc}
\toprule
\multirow{2}{*}{\textbf{Topology}} &
\multirow{2}{*}{\textbf{Framework}} &
\multicolumn{3}{c}{\textbf{Baseline}} &
\multicolumn{3}{c}{\textbf{Compliance}} &
\multicolumn{3}{c}{\textbf{Security FUD}} \\
\cmidrule(lr){3-5} \cmidrule(lr){6-8} \cmidrule(lr){9-11}
& & \textbf{M} & \textbf{Q} & \textbf{R}
  & \textbf{M} & \textbf{Q} & \textbf{R}
  & \textbf{M} & \textbf{Q} & \textbf{R} \\
\midrule

\multirow{2}{*}{Chain}
& LangChain
& \cellcolor{goodgreen}3.3  & \cellcolor{goodgreen}0.0  & \cellcolor{goodgreen}0.0
& \cellcolor{midorange}95.0 & \cellcolor{midorange}96.7 & \cellcolor{midorange}85.0
& \cellcolor{badred}100.0 & \cellcolor{badred}100.0 & \cellcolor{badred}100.0 \\
& MetaGPT
& 5.0  & 11.7 & \cellcolor{midorange}46.7
& \cellcolor{badred}100.0 & \cellcolor{midorange}98.3 & \cellcolor{badred}100.0
& \cellcolor{midorange}96.7 & 76.7 & \cellcolor{midorange}95.0 \\

\midrule

\multirow{2}{*}{Mesh}
& AutoGen
& \cellcolor{goodgreen}0.0  & 5.0  & \cellcolor{goodgreen}0.0
& \cellcolor{badred}100.0 & \cellcolor{badred}100.0 & \cellcolor{midorange}95.0
& \cellcolor{midorange}98.3 & \cellcolor{badred}100.0 & \cellcolor{midorange}98.3 \\
& CAMEL
& \cellcolor{goodgreen}0.0  & \cellcolor{goodgreen}0.0  & \cellcolor{goodgreen}0.0
& \cellcolor{badred}100.0 & \cellcolor{badred}100.0 & \cellcolor{badred}100.0
& \cellcolor{badred}100.0 & \cellcolor{midorange}98.3 & \cellcolor{badred}100.0 \\

\midrule

\multirow{2}{*}{Star}
& CrewAI
& \cellcolor{goodgreen}0.0  & \cellcolor{goodgreen}0.0  & \cellcolor{goodgreen}3.3
& 46.7 & 51.7 & 31.7
& 46.7 & 43.3 & 33.3 \\
& LangGraph
& \cellcolor{goodgreen}0.0  & 8.3  & 11.7
& \cellcolor{badred}100.0 & \cellcolor{badred}100.0 & \cellcolor{badred}100.0
& \cellcolor{badred}100.0 & \cellcolor{midorange}98.3 & \cellcolor{badred}100.0 \\

\bottomrule
\end{tabular*}
\end{table*}

\noindent{\textbf{Tasks and scenarios.}}
We evaluate three scenarios covering code-centric collaboration, strict rule-following, and general knowledge question answering:
\textsc{Quant}, consisting of data analysis tasks constructed from the UCI repository~\cite{uci};
\textsc{Rigid}, comprising multi-step logic and calculation problems derived from MATH benchmarks~\cite{math};
and \textsc{MMLU}, adapting general knowledge questions from the MMLU benchmark~\cite{mmlu} into retrieval-based tasks.

\vspace{2pt}

\noindent \textbf{Frameworks and topologies.}
We test six mainstream MAS frameworks spanning three interaction topologies. The chain topology includes \textsc{LangChain}\cite{langchain} and \textsc{MetaGPT}\cite{metagpt}. The mesh topology includes \textsc{AutoGen}\cite{wu2024autogen} and \textsc{CAMEL}\cite{li2023camel}. The star topology includes \textsc{CrewAI}\cite{crewai} and \textsc{LangGraph}\cite{langgraph}.

\vspace{2pt}

\noindent \textbf{Attack settings.}
We instantiate three attacker policies via \textbf{\textit{application-level}} message injection, in contrast to the \textbf{\textit{system-level}} injection used in Section~\ref{sec:vuln:cascade}: \textsc{Baseline} (direct insertion of the raw malicious claim), \textsc{Compliance}, and \textsc{Security\_FUD}. For each task instance, we measure whether a single injected seed can be propagated through multi-round collaboration into an infected final artifact.

\vspace{2pt}

\noindent \textbf{Defense settings.}
We evaluate \textsc{None}, three operating points of our governance layer, and three comparison baselines.
We use self-reflection as an agent-side self-check baseline, where the agent reviews and revises its own output before sending it~\cite{ji2023towards}.
We adapt AGrail as a runtime guardrail baseline that checks each outgoing inter-agent message before it enters downstream context~\cite{luo2025agrail}.
We further include a control-flow guard (CFG) baseline inspired by ControlValve~\cite{jha2025breaking}, which checks runtime transitions against admissible edge rules and attempts an alternative valid route when a transition is rejected.
AGrail and CFG are evaluated at their native granularity. AGrail operates on whole inter-agent messages, CFG operates on routing transitions, and claim-lineage rollback is evaluated as part of our approach.
To simplify the experimental design, we select one framework per topology for the defense experiments.
For atomic decomposition, we adopt the FActScore protocol and employ GPT-4o-mini to decompose agent messages~\cite{min2023factscore}.
For entailment, we use a natural language inference (NLI) model based on DeBERTa-v3-small~\cite{he2020deberta}.
We report a supplementary evaluation of the NLI module on a human-annotated atomic-fact set derived from our real tasks in Appendix~\ref{app:appendix-c}.
For the verification layer, we use a self-built knowledge base and GPT-4o-mini.

\vspace{2pt}

\noindent \textbf{Metrics and observables.}
We report ASR, BICR, Safe Completion, Token/Safe, and Latency/Safe.
Attack Success Rate (ASR) is the fraction of tasks whose final artifact is labeled infected.
Benign Infection Control Rate (BICR) is defined as $1 - \mathrm{ASR}$.
BICR measures whether the final artifact avoids infection. We additionally report Safe Completion, defined as the fraction of tasks that produce a usable and non-infected final artifact.
Token/Safe is the average total token consumption over safe completions.
Latency/Safe is the corresponding average runtime in seconds.
Infection labels are computed by a deterministic evaluation harness.
In addition to final outcomes, we log per-round infection states and the trajectory-level infection coverage $S(t)$ to visualize whether corruption remains local or escalates system-wide.

\vspace{2pt}

\noindent \textbf{Reproducibility details.}
All evaluation details (including datasets, framework configurations, attack prompts, defense configuration, NLI module evaluation, and labeling rules) are provided in Appendix~\ref{app:appendix-c}.

\subsection{Adversarial Severity}
\label{sec:eval-attack}

We quantify adversarial severity at the application layer. Table~\ref{tab:asr_compact} reports ASR for six frameworks across three scenarios and policies. Intent-hiding attackers consistently outperform direct injection. While \textsc{Baseline} is near zero in almost all settings and only rises on \textsc{MetaGPT}--\textsc{Rigid} (46.7\%), \textsc{Compliance} pushes ASR to mostly 85.0--100.0\% outside \textsc{CrewAI}, and \textsc{Security FUD} further raises ASR to 76.7--100.0\% in those settings, indicating that role-consistent packaging substantially increases end-to-end corruption probability over plain injection. The weakest intent-hiding point outside \textsc{CrewAI} appears on \textsc{MetaGPT}--\textsc{Quant} under \textsc{Security FUD} (76.7\%), suggesting that scenario constraints and workflow choices partially limit adoption.

Cross-framework differences remain visible under the same topology label. In mesh-style collaboration, both \textsc{AutoGen} and \textsc{Camel} saturate under intent-hiding attacks, reaching up to 100.0\% ASR across scenarios. In star-style orchestration, \textsc{LangGraph} similarly reaches 100.0\% ASR under \textsc{Compliance} and almost saturates under \textsc{Security FUD}, whereas \textsc{CrewAI} is substantially lower. This gap suggests implementation-level coordination and routing policies materially affect how injected content is reused and endorsed, beyond coarse topology classification.

\begin{table*}[t]
\centering
\small
\setlength{\tabcolsep}{7pt}
\renewcommand{\arraystretch}{1.12}

\caption{BICR, safe completion, and token cost across frameworks and defenses.}
\label{tab:master_all}

\begin{tabular*}{\textwidth}{@{\extracolsep{\fill}} ll ccc ccc ccc}
\toprule
\multirow{2}{*}{\textbf{Framework}} &
\multirow{2}{*}{\textbf{Defense}} &
\multicolumn{3}{c}{\cellcolor{groupgray}\textbf{BICR (\%) $\uparrow$}} &
\multicolumn{3}{c}{\cellcolor{groupgray}\textbf{Safe Completion (\%) $\uparrow$}} &
\multicolumn{3}{c}{\cellcolor{groupgray}\textbf{Token / Safe ($10^3$) $\downarrow$}} \\
\cmidrule(lr){3-5}\cmidrule(lr){6-8}\cmidrule(lr){9-11}
& & \textbf{Base.} & \textbf{Comp.} & \textbf{Sec.}
  & \textbf{Base.} & \textbf{Comp.} & \textbf{Sec.}
  & \textbf{Base.} & \textbf{Comp.} & \textbf{Sec.} \\
\midrule

\multirow{6}{*}{\textbf{MetaGPT}}
& Strict
& 100.0 & 95.0 & 100.0
& \cellcolor{goodgreen}95.0 & \cellcolor{lightgreen}89.2 & \cellcolor{goodgreen}97.5
& 52.9 & 49.6 & 52.0 \\
& Balanced
& 97.5 & 88.3 & 100.0
& \cellcolor{goodgreen}93.3 & \cellcolor{lightgreen}88.3 & \cellcolor{goodgreen}92.5
& 24.3 & 23.2 & 24.0 \\
& Speed
& 95.8 & 85.0 & 100.0
& \cellcolor{goodgreen}91.7 & \cellcolor{lightgreen}79.2 & \cellcolor{goodgreen}97.5
& 17.1 & 16.0 & 16.5 \\
& Reflection
& 82.5 & 5.0 & 19.2
& \cellcolor{lightgreen}80.0 & \cellcolor{badred}5.0 & \cellcolor{badred}19.2
& 9.7 & 9.3 & 8.7 \\
& AGrail
& 96.7 & 36.7 & 42.5
& \cellcolor{midorange}40.8 & \cellcolor{badred}0.8 & \cellcolor{badred}2.5
& 14.2 & 14.3 & 15.2 \\
& CFG
& 80.8 & 40.0 & 40.8
& \cellcolor{midorange}53.3 & \cellcolor{badred}0.0 & \cellcolor{badred}3.3
& 6.7 & -- & 7.1 \\
\midrule

\multirow{6}{*}{\textbf{AutoGen}}
& Strict
& 99.2 & 100.0 & 100.0
& \cellcolor{goodgreen}97.5 & \cellcolor{goodgreen}100.0 & \cellcolor{goodgreen}100.0
& 75.8 & 77.8 & 84.9 \\
& Balanced
& 100.0 & 95.0 & 96.7
& \cellcolor{goodgreen}100.0 & \cellcolor{goodgreen}92.5 & \cellcolor{goodgreen}95.8
& 29.2 & 30.1 & 33.0 \\
& Speed
& 100.0 & 96.7 & 98.3
& \cellcolor{goodgreen}99.2 & \cellcolor{goodgreen}96.7 & \cellcolor{goodgreen}98.3
& 29.2 & 30.0 & 32.9 \\
& Reflection
& 100.0 & 5.8 & 3.3
& \cellcolor{goodgreen}96.7 & \cellcolor{badred}5.8 & \cellcolor{badred}3.3
& 18.0 & 17.4 & 19.4 \\
& AGrail
& 100.0 & 89.2 & 79.2
& \cellcolor{badred}8.3 & \cellcolor{badred}0.0 & \cellcolor{badred}3.3
& 28.4 & -- & 30.3 \\
& CFG
& 99.2 & 84.2 & 75.8
& \cellcolor{midorange}35.0 & \cellcolor{badred}0.8 & \cellcolor{badred}5.8
& 26.8 & 29.1 & 27.1 \\
\midrule

\multirow{6}{*}{\textbf{LangGraph}}
& Strict
& 92.5 & 85.0 & 76.7
& \cellcolor{goodgreen}92.5 & \cellcolor{lightgreen}85.0 & \cellcolor{lightgreen}76.7
& 37.1 & 37.9 & 41.5 \\
& Balanced
& 95.8 & 85.8 & 76.7
& \cellcolor{goodgreen}95.8 & \cellcolor{lightgreen}85.8 & \cellcolor{lightgreen}76.7
& 35.9 & 38.3 & 41.5 \\
& Speed
& 94.2 & 58.3 & 76.7
& \cellcolor{goodgreen}94.2 & \cellcolor{midorange}58.3 & \cellcolor{lightgreen}76.7
& 13.6 & 14.4 & 15.6 \\
& Reflection
& 73.3 & 0.8 & 0.8
& \cellcolor{lightgreen}73.3 & \cellcolor{badred}0.8 & \cellcolor{badred}0.8
& 9.8 & 11.0 & 13.4 \\
& AGrail
& 95.8 & 90.0 & 79.2
& \cellcolor{midorange}44.2 & \cellcolor{badred}0.0 & \cellcolor{badred}0.0
& 23.2 & -- & -- \\
& CFG
& 98.3 & 90.0 & 77.5
& \cellcolor{midorange}48.3 & \cellcolor{badred}0.0 & \cellcolor{badred}0.0
& 9.4 & -- & -- \\
\bottomrule
\end{tabular*}

\vspace{4pt}
\begin{minipage}{1\textwidth}
\footnotesize
Base., Comp., and Sec. refer to Baseline, Compliance, and Security FUD settings. “–” means no safe completion was produced under corresponding setting.
\end{minipage}

\end{table*}

\subsection{Defense Efficacy of the Governance Layer}
\label{sec:eval-defense}

We evaluate whether the governance plug-in can contain propagation under immutable topology and preserved agent availability, by intervening only at the message layer. Table~\ref{tab:master_all} reports BICR, Safe Completion, and Token/Safe across attack settings and defense methods. The joint view is necessary because a method may obtain a non-infected output by interrupting the workflow, so Safe Completion is used to check whether the output remains usable. Figure~\ref{fig:defense_trajectories} further shows that Speed, Balanced, and Strict flatten $S(t)$ earlier than \textsc{None} across all three topologies.

Across MetaGPT and AutoGen, the proposed modes generally maintain high BICR and Safe Completion under intent-hiding attacks. On MetaGPT, Speed reaches 85.0\% BICR under Compliance and 100.0\% under Security FUD, with Safe Completion at 79.2\% and 97.5\%. On AutoGen, all three proposed modes keep BICR above 95.0\% under Compliance and Security FUD, and Safe Completion remains above 92.5\%. LangGraph is less stable, especially under Security FUD, where Strict and Balanced both obtain 76.7\% BICR. Even in this setting, the proposed modes still provide higher usable non-infected outputs than the agent-monitoring and graph-control baselines in most cases.

Reflection preserves the original workflow but remains weak against intent-hiding messages, with BICR falling to 5.0\%/19.2\% on MetaGPT, 5.8\%/3.3\% on AutoGen, and 0.8\%/0.8\% on LangGraph under Compliance and Security FUD. AGrail and CFG show a different pattern: they can reduce final infection in some settings, but Safe Completion remains limited. Under Compliance, AutoGen reports 89.2\%/0.0\% for AGrail and 84.2\%/0.8\% for CFG in BICR/Safe Completion, while LangGraph reports 90.0\%/0.0\% for both.

The separation between BICR and Safe Completion suggests that final infection control should be interpreted together with usable task completion. Agent-monitoring and graph-control baselines can stop part of the corrupted propagation, but they often reduce the chance of producing a usable final artifact. By removing unsafe atomic claims through screening and rollback, the governance plug-in limits propagation while keeping the downstream workflow usable in more attack settings.


\vspace{2pt}

\noindent \textbf{Ablation study.} 
To isolate the contribution of each layer, we remove atomization (\texttt{no\_atomization}), screening and verification (\texttt{no\_detection}), and quarantine or rollback (\texttt{no\_blocking}); Table~\ref{tab:ablation_layered} reports the corresponding overheads under the Strict policy. The ablation results show a clear drop in containment once key components are removed. Without atomization, BICR remains at $40.0 \pm 49.0\%$, which is higher than the other ablated variants but also highly variable across runs. Removing screening and verification lowers BICR to $14.4 \pm 35.2\%$, indicating that explicit risk decisions are a major contributor to containment. Removing quarantine or rollback further reduces BICR to $3.1 \pm 10.1\%$, close to the \textsc{None} setting at $2.2 \pm 14.7\%$.

The overhead pattern helps explain the role of enforcement. The \texttt{no\_blocking} variant consumes the most tokens, $34{,}991$, and takes $85.7 \pm 21.0$s, yet its BICR is only $3.1\%$. This suggests that screening without an effective quarantine or rollback step can still spend substantial computation while leaving propagation largely uncontrolled. In contrast, \texttt{no\_atomization} uses fewer tokens, $11{,}663$, and achieves the highest BICR among the ablations, but the large variance prevents a stronger stability claim. Overall, the results suggest that detection signals need enforceable isolation or rollback to affect downstream propagation, while atomization and screening help make this enforcement more targeted.


\begin{table}[t]
\centering
\normalsize
\caption{Ablation study results.}
\label{tab:ablation_layered}
\vspace{3pt}
\setlength{\tabcolsep}{5pt}
\renewcommand{\arraystretch}{1.00}

\begin{tabular*}{0.96\columnwidth}{@{\extracolsep{\fill}}lccc@{}}
\toprule
\textbf{Variant} & \textbf{BICR (\%) } & \textbf{Tokens} & \textbf{Time (s) } \\
\midrule
w/o Atomization & $40.0 \pm 49.0$ & 11{,}663 & $77.9 \pm 24.7$ \\
w/o Detection   & $14.4 \pm 35.2$ & 14{,}708 & $51.7 \pm 13.0$ \\
w/o Blocking    & $3.1 \pm 10.1$  & 34{,}991 & $85.7 \pm 21.0$ \\
None            & $2.2 \pm 14.7$  & 7{,}365  & $41.2 \pm 11.9$ \\
\bottomrule
\end{tabular*}
\end{table}

\subsection{The Cost of Safety}
\label{sec:eval-cost}

We measure overhead with Token/Safe and Latency/Safe, which report token use and runtime over safe completions. Table~\ref{tab:cost_summary} summarizes the averaged safety-cost trade-off across proposed modes and comparison baselines. Moving from Reflection to Speed gives the largest robustness gain: BICR rises from 0.32 to 0.89 and Safe Completion from 0.32 to 0.88, while Token/Safe increases from 12,749 to 21,227 and Latency/Safe from $91.8 \pm 24.0$s to $149.7 \pm 39.5$s.

Balanced and Strict further reduce residual infection, with BICR increasing from 0.89 under Speed to 0.93 and 0.94. Safe Completion also rises from 0.88 to 0.91 and 0.93. The additional robustness requires a clear cost increase: Token/Safe grows from 21,227 under Speed to 30,844 under Balanced and 57,610 under Strict, while Latency/Safe increases from $149.7 \pm 39.5$s to $179.6 \pm 44.3$s and $217.9 \pm 53.7$s. This trend supports using Speed as the cost-aware mode and Strict as the high-assurance mode when the application can tolerate higher latency and token usage.

AGrail and CFG have lower Latency/Safe than the proposed modes, at $98.8 \pm 20.5$s and $65.3 \pm 25.2$s, respectively. However, their Safe Completion remains low, at 0.11 and 0.16, despite BICR values of 0.79 and 0.76. This gap indicates that lower runtime does not necessarily yield a better deployment trade-off. Agent-monitoring and graph-control baselines can reduce final infection in some runs, but still produce limited usable non-infected outputs. The proposed modes require more tokens and time, yet preserve a larger fraction of safe completed workflows.



\begin{table}[t]
\centering
\normalsize
\caption{Safety-cost summary.}
\label{tab:cost_summary}
\vspace{3pt}
\setlength{\tabcolsep}{4.5pt}
\renewcommand{\arraystretch}{1.12}

\begin{tabular*}{1\columnwidth}{@{\extracolsep{\fill}}lcccc@{}}
\toprule
\textbf{Mode} & \textbf{BICR} & \textbf{Safe Comp.} & \textbf{Tok./Safe} & \textbf{Lat./Safe (s)} \\
\midrule
Reflection & 0.32 & 0.32 & 12{,}749 & $91.8 \pm 24.0$ \\
Speed      & 0.89 & 0.88 & 21{,}227 & $149.7 \pm 39.5$ \\
Balanced   & 0.93 & 0.91 & 30{,}844 & $179.6 \pm 44.3$ \\
Strict     & 0.94 & 0.93 & 57{,}610 & $217.9 \pm 53.7$ \\
AGrail     & 0.79 & 0.11 & 19{,}902 & $98.8 \pm 20.5$ \\
CFG        & 0.76 & 0.16 & 13{,}323 & $65.3 \pm 25.2$ \\
\bottomrule
\end{tabular*}


\end{table}

\section{Limitations and Future Work}
\label{sec:limitations}

We isolate an LLM-MAS failure mechanism where local errors propagate through message dependencies into an incorrect final artifact. We contribute a propagation abstraction with trajectory observables such as $S(t)$, evidence of endogenous vulnerabilities across major frameworks, and a message-layer governance plug-in requiring no retraining. The model is defined over continuous states $s_i(t)\in[0,1]$ and aggregate trajectories $S(t)$, while evaluation uses a binary variable $X_i(t)\in\{0,1\}$ as a simplified interface. This improves interpretability and comparability but may miss cases where agents partially internalize incorrect premises, relax constraints, or show semantic drift without meeting strict adoption criteria. The abstraction treats $(\beta,\delta)$ and structural term $A$ as stationary summaries and captures role, task, or time variation only indirectly. Hence, the spectral indicator $\mathcal{R}$ is a theoretical heuristic for early analysis and evidence organization, not a deployment guarantee or calibrated risk predictor.

The injection study is a controlled probe of amplification, not a full characterization of adversarial behavior. We seed application-layer messages to fix an initial condition and test whether endogenous amplification propagates a small seed to the final outcome. Here, the attacker lacks persistent system access and execution-layer control, and the governance plug-in operates at the same interface to suppress downstream adoption. This isolates propagation effects from stronger capabilities such as tool tampering and long-term profile poisoning; multi-stage adaptive adversaries that react to intermediate states, or have richer access, remain out of scope and are a natural extension.

The governance plug-in incurs latency overhead in online mode, reflecting the trade-off between runtime safety and execution speed. It also maintains genealogical traces for offline auditing, diagnosis, and post hoc analysis. Our defense evaluation measures safety gains under online deployment, with screening and verification integrated into the workflow. Synchronous execution on the interaction path can increase latency; decoupling or partial offloading shifts protection to non-critical analysis. Balancing these configurations to meet real-time latency budgets and safety targets remains a deployment constraint, not a resolved systems issue.


\noindent \textbf{Future Work.}
Future work will strengthen validation and broaden the settings where this propagation view and governance logic apply. One direction studies longer-horizon tool-using workflows where failures emerge as downstream artifacts, decisions, or plans, testing whether similar amplification signatures appear in software engineering and scientific support. A second direction refines measurement by improving adoption ground truth beyond binary observables and tightening estimation of structural quantities, including $\rho(A)$, from real traces. A third direction refines system integration by coupling message provenance, evidence retrieval, and verification scheduling, jointly tuning online and offline use to improve the safety-latency trade-off in realistic deployments.


\section{Related Work}

\noindent \textbf{Adversarial Attacks on LLM-MAS.}
Integrating LLMs into multi-agent workflows expands the attack surface across user prompts, inter-agent messages, and tool interfaces; surveys catalog prompt injection, context hijacking, and protocol-level manipulation of agent communication \cite{threats2025workflow,kong2025survey}. In multi-agent settings, malicious outputs can propagate through downstream dependencies: indirect prompt injection can traverse multi-turn delegations, override system intent, and amplify impact beyond the originating agent \cite{sentinel2025graph,wang2025security}. Attackers can also bias group decisions by steering one agent or the coordination channel, turning cooperative verification into an amplification path and exposing system-spanning weaknesses uncovered by model-local safeguards \cite{wang2025large,acl2025consensus}. Recent exploits demonstrate this systemic gap, including Agent-in-the-Middle interception of inter-agent messages, and the web, tool, and shared memory combination allowing one agent's unsafe intake to trigger harmful actions in another, with reports of MAS executing attacker-supplied code at high success rates \cite{acl2025consensus,lupinacci2025dark,d2021nowhere}. These observations motivate orchestration-layer governance, including trust policies and secure communication design rather than ad hoc per-agent defenses \cite{shmatikov2025arbitrary,zeydan2023blockchain,adimulam2026orchestration}.

\vspace{2pt}

\noindent \textbf{Failure Modes of LLM-MAS}
Even without adversaries, LLM multi-agent systems exhibit intrinsic failure modes, and empirical work finds chaining agents often yields limited reliability gains while introducing multi agent specific breakdowns \cite{pan2025multiagent}. Common symptoms include role misunderstanding, agents overwriting each other's context, and stalling in unproductive loops, compounded by memory and context drift in long interactions where earlier constraints are forgotten and critical details are dropped \cite{pan2025multiagent,sapkota2025ai,sagallm2025context}. Without external verification, agents can reinforce shared errors, creating echo effects that inflate confidence in wrong conclusions, and small inaccuracies accumulate over repeated cycles when no corrective mechanism intervenes \cite{baker2025monitoring,acl2025consensus}. A large-scale analysis of more than 1,600 MAS traces systematizes these issues into 14 failure modes across System Design Issues, Inter Agent Misalignment, and Task Verification Failures, highlighting weak delegation and ambiguous instructions as recurring roots \cite{pan2025multiagent}. SagaLLM further illustrates missing rollback and global state checks when parallel bookings conflict after a flight cancellation, reflecting absent coordination and state management principles standard in distributed systems and leaving architectures unable to guarantee consistency, recovery, and cross-agent transparency \cite{sagallm2025context}.

\vspace{2pt}


\noindent \textbf{Defenses and Detectors for LLM-MAS}
To mitigate risks, defenses span per-agent guardrails and system-level monitoring. Structured prompts, output filters, and policy tools constrain individual behavior but operate locally \cite{adabara2025trustworthy}. Since isolated protections lack visibility into cross-agent dependencies, they miss distributed attack sequences and workflow inconsistencies arising after messages and tool calls traverse team \cite{sentinel2025graph,ray2025review,foundjem2025multi}. System-level approaches add redundancy and cross-checking, introducing validator agents, watchdog frameworks, and graph-based anomaly detection over interaction patterns, e.g., SentinelAgent's interaction graph plus LLM monitoring and log audits like Audit-LLM for insider threats \cite{zoppi2017exploring,acl2025consensus,song2024audit}. However, inter-agent links remain under-protected; multi-hop attacks evade detectors, adversarial steering subverts consensus reliability, and monitoring overhead limits deployment \cite{aslam2022overview,shmatikov2025arbitrary,ishii2022overview}. A consistent direction places security controls in the orchestration layer via trust policies, unified logging, supervisory halt or rollback, least-privilege access, and communication provenance, enabling global consistency through mechanisms like checkpointing or transactional memory while limiting adversarial influence \cite{ray2025review,islam2019multi}.

\section{Conclusion}
We show that a single atomic falsehood can propagate and stabilize as system-level false consensus in fixed LLM-MAS workflows, and we make this process measurable by fitting discrete-time IBMF dynamics on the communication graph with $s_i(t)\in[0,1]$ and $S(t)$.  The fitted interface separates transmission, driven by $\beta$ and $\rho(A)$, from correction, driven by $\delta$, yielding the early amplification condition $\beta\rho(A)>\delta$ and the auxiliary risk indicator $R$. 
In evaluation across multiple mainstream frameworks, topologies, and task scenarios, plain injection is near zero in most settings, whereas role-consistent intent-hiding packaging sharply increases end-to-end corruption; the same lens also explains strong sensitivity to structurally central injection points and the decreasing reversibility of late corrections as polluted context accumulates. Finally, we introduce a genealogy-based governance layer that keeps $A$ unchanged and intervenes only on the message path via claim provenance, tri-state screening, targeted verification, and enforced rollback; results and ablations show strong containment, and further indicate that detection alone does not reliably contain propagation without rollback or isolation.    
\cleardoublepage

\appendices
\label{sec:app}

\section*{Ethical Considerations}
This paper studies how small local errors can propagate through collaboration in LLM-based multi-agent systems and lead to a shared but incorrect outcome, and it evaluates mitigation mechanisms under controlled experimental settings.

\vspace{2pt}

\noindent \textbf{Stakeholders.} Potentially impacted stakeholders include (1) developers and operators of LLM-MAS frameworks, who may face higher risk when inter-agent message exchange is manipulated or adversarial content is injected into collaborative workflows. Our mitigation mechanisms aim to reduce this risk by making unsupported claims harder to spread and easier to correct during collaboration. (2) End users of applications built on LLM-MAS may be affected when collaborative outputs drift from verifiable evidence, leading to reduced correctness or reliability in the final artifact. They also benefit when the system can surface unsupported claims early and support timely correction before the final output is produced. (3) security practitioners and researchers, who benefit from evidence about when and why collaboration fails; and (4) the broader online ecosystem, where collaborative tooling can scale both beneficial and harmful content.

\vspace{2pt}

\noindent \textbf{Potential impacts.}
The primary risk is that the attack principles studied here could be adapted to bias collaborative systems by packaging a false seed as credible and encouraging repeated reuse across agents. At the same time, the core technical components studied in this work support responsible mitigation. In particular, atomic claim decomposition and structured intervention mechanisms can improve transparency during collaboration, support post hoc analysis of how an incorrect belief was adopted, and strengthen accountability by making correction points explicit. We believe the benefits outweigh the risks when disclosure is bounded and safeguards are applied.

\vspace{2pt}

\noindent \textbf{Risk boundaries and safeguards.}
Our attack experiments are conducted only in controlled evaluation settings and do not target production systems or real user traffic. The injected content includes controlled, fictitious security-themed messages that are intentionally synthetic and are not claims about real vulnerabilities in real infrastructure. We do not provide any one-click automated hacking toolkit or deployment-ready abuse pipeline. We release only attack principles and the experimental scripts needed to reproduce the paper results, and we avoid packaging guidance that would materially reduce the barrier for misuse.

\vspace{2pt}

\noindent \textbf{Human subjects and annotation.}
When human annotation is used, annotators participate voluntarily and can stop at any time. The annotation content does not involve sensitive personal information, and we report only aggregated results.

\vspace{2pt}

\noindent \textbf{Data and privacy.}
The released datasets are sanitized and do not contain real personal data. Any URLs, database identifiers, or similar fields appearing in artifacts are synthetic.

\section*{Open Science}
To support reproducibility and NDSS artifact evaluation, we release the artifacts needed to evaluate our contribution via an anonymized repository link: \url{https://anonymous.4open.science/r/From-spark-to-fire-6E0C/}. The release includes source code, datasets, and experiment scripts sufficient to reproduce the reported results. To reduce dual-use risk, the artifact package does not include any deployment-ready automation intended for real-world misuse.

\bibliographystyle{IEEEtran}
\bibliography{mybibliography}

\section{Model Fitting and Topology Configuration Details}
\label{app:fit-topology}

This appendix specifies the configuration and fitting protocol omitted from \ref{modelutility}. In this calibration experiment, we operationalize ``infection'' using a neutral tracer: a randomly generated codeword that carries no task semantics. A node is marked infected once its output contains the injected codeword. This design avoids coupling the seed to correctness and provides a clean signal for validating whether the fitted dynamics and the communication topology match the observed adoption trajectories.

\vspace{2pt}

\noindent \textbf{Setup and topology definitions} We fix the network size to $N=5$, the horizon to $T=5$ synchronous interaction rounds, and the number of independent trials to $R=20$ for each topology. Let $G=(V,E)$ be a directed graph with $V=\{0,1,2,3,4\}$. A directed edge $(j,i) \in E$ means agent $i$ can read messages produced by agent $j$ in that round. The adjacency matrix $A \in \{0,1\}^{N\times N}$ is defined by $a_{ij}=1$ if $(j,i)\in E$ and $a_{ij}=0$ otherwise.

\begin{figure}[t]
    \centering
    \includegraphics[width=\linewidth]{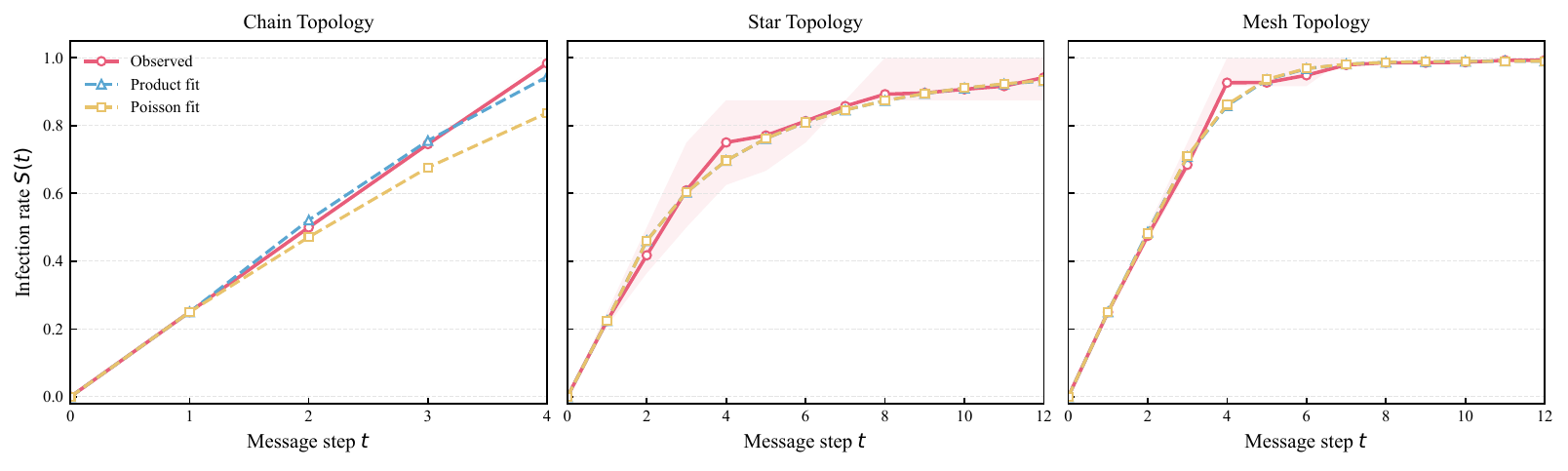}
    \caption{\textbf{Real attack trace fitting under asynchronous message steps.} The observed curve gives the mean infection rate $S(t)$ over real attack traces grouped by topology. The shaded band is the central $50\%$ run-level interval of the observed trajectories, computed from the 25th and 75th percentiles.}
    \label{fig:real_attack_fit}
\end{figure}

The centralized topology is a bidirectional star with node $0$ as the hub:
\[
E_{\mathrm{star}}=\{(0,i)\mid i\in\{1,\dots,4\}\}\cup\{(i,0)\mid i\in\{1,\dots,4\}\}.
\]
The layered topology is a unidirectional chain,
\[
E_{\mathrm{chain}}=\{(i,i+1)\mid i\in\{0,1,2,3\}\}.
\]
The layered-horizontal topology augments this chain with static reverse edges that represent limited two-way communication: for each adjacent pair $(i,i+1)$, we draw $B_i\sim\mathrm{Bernoulli}(P_h)$ once at the start of the experiment with $P_h=0.3$; if $B_i=1$, we add the reverse edge $(i+1,i)$ and obtain
\[
E_{\mathrm{lh}}=E_{\mathrm{chain}}\cup\{(i+1,i)\mid i\in\{0,1,2,3\},\ B_i=1\}.
\]
We do not generate skip connections; the jump probability is fixed to $P_s=0$. The decentralized topology is implemented as the complete directed graph without self-loops,
\[
E_{\mathrm{complete}}=\{(j,i)\mid i\neq j,\ i,j\in V\}.
\]

\noindent \textbf{Codeword matching and tolerance rules} In the modeling assessment part, at the beginning of each trial we sample a fresh codeword string $c$ in a sentinel-delimited format (for example, \texttt{\#TAG:ABC-1234\#}) and inject it at the seed position.  Matching is literal and case sensitive. Before detection, each agent output $y_{i,r}(t)$ is passed through a deterministic normalization function $\mathrm{norm}(\cdot)$ that applies Unicode normalization (NFKC), standardizes line endings, removes zero-width characters, and strips a single layer of trivial wrappers such as surrounding backticks or quotes around $c$. We then mark agent $i$ as infected at round $t$ in trial $r$ if and only if $c$ appears as a substring of $\mathrm{norm}(y_{i,r}(t))$. We count only exact matches of the full codeword and do not accept partial matches or edited variants; inserted whitespace, case changes, dropped delimiters, or substituted characters are all treated as non-infected.

\vspace{2pt}

\noindent \textbf{Observed trajectory and fitting protocol} Let $X_{i,r}(t)\in\{0,1\}$ denote whether agent $i$ is marked infected at round $t$ in trial $r$ under the codeword rule above. The observed system-level infection rate is
\[
S_{\mathrm{obs}}(t)=\frac{1}{RN}\sum_{r=1}^{R}\sum_{i=0}^{N-1} X_{i,r}(t), \qquad t\in\{1,\dots,T\}.
\]
If a trial reaches full infection at time $t_{\mathrm{stop}}<T$ (that is, $X_{i,r}(t_{\mathrm{stop}})=1$ for all $i$), we set $X_{i,r}(t)=1$ for all $t\in(t_{\mathrm{stop}},T]$ so that steady-state segments are not truncated.

The update rule and infection functions follow \ref{modelutility}. Since round $t=0$ contains only the seed injection, we align the model initial condition to the first observed interaction round by setting a homogeneous initialization
\[
s_i^{\mathrm{model}}(0)\leftarrow S_{\mathrm{obs}}(1)\quad\text{for all } i.
\]
The model then produces a predicted aggregate trajectory $S_{\mathrm{pred}}(t)=\frac{1}{N}\sum_i s_i^{\mathrm{model}}(t)$, which we compare to $S_{\mathrm{obs}}(t+1)$ for $t=1,\dots,T-1$. The mean squared error is
\[
\mathrm{MSE}(\beta,\delta)=\frac{1}{T-1}\sum_{t=1}^{T-1}\Bigl(S_{\mathrm{pred}}(t)-S_{\mathrm{obs}}(t+1)\Bigr)^2.
\]

We select $(\beta^*,\delta^*)$ through a two-stage grid search. The coarse stage evaluates $(\beta,\delta)\in[0,1]\times[0,1]$ with step size $0.05$. The fine stage evaluates an $11\times 11$ grid centered at the coarse optimum over a radius of $\pm 0.05$ with step size $0.01$, clipping candidates to $[0,1]$ at the boundaries.

\vspace{2pt}

\noindent \textbf{Fitting on Real Attack Traces} To further examine the applicability of the model in real attack settings, we conduct an offline fitting analysis on the real attack logs used in the application-level attack experiment. Compared with the neutral-codeword experiment, the propagated signal in real attack traces is embedded in concrete malicious instructions and task contexts. Therefore, this experiment serves as a real-scenario extension for evaluating the model's fitting ability.

\begin{table}[t]
\centering
\caption{Fit on real attack traces.}
\label{tab:real_attack_fit}
\normalsize
\setlength{\tabcolsep}{4.8pt}
\renewcommand{\arraystretch}{1.12}
\begin{tabular}{@{}llrrrr@{}}
\toprule
\textbf{Framework} & \textbf{Topology} & $\boldsymbol{\beta}$ & $\boldsymbol{\delta}$ & \textbf{MSE} & \textbf{Final ASR} \\
\midrule
AutoGen   & Mesh  & 0.363 & 0.002 & 0.0274 & 0.986 \\
CAMEL     & Mesh  & 0.270 & 0.010 & 0.0216 & 0.997 \\
CrewAI    & Star  & 0.220 & 0.020 & 0.0471 & 0.422 \\
LangChain & Chain & 0.985 & 0.002 & 0.0001 & 0.961 \\
LangGraph & Star  & 0.470 & 0.000 & 0.0310 & 0.997 \\
MetaGPT   & Chain & 0.593 & 0.000 & 0.0468 & 0.944 \\
\bottomrule
\end{tabular}
\end{table}

The product-form fit gives a mean MSE of $0.029$ on the pooled attack groups. This result shows that the propagation model can also fit the coarse-grained adoption curves in real attack traces.

\noindent \textbf{Compact threshold diagnostics.}
To complement the representative threshold-shift analysis, we further report compact fitted diagnostics on two representative governance settings, LangGraph-Star and AutoGen-Mesh. Since the ratio $R=\beta\rho(A)/\delta$ becomes ill-conditioned when $\delta$ is close to zero, we report the threshold margin $\beta\rho(A)-\delta$ as the primary diagnostic.

\begin{table}[t]
\centering
\caption{Compact propagation-threshold diagnostics.}
\label{tab:compact_threshold_diagnostics}
\small
\setlength{\tabcolsep}{4.5pt}
\renewcommand{\arraystretch}{1.08}
\begin{tabular}{llcccc}
\toprule
\textbf{Setting} & \textbf{Defense} & $\beta$ & $\delta$ & \textbf{Margin} & Obs. $S(T)$ \\
\midrule
LangGraph-Star & None   & 0.42 & 0.08 & 0.65 & 0.80 \\
LangGraph-Star & Speed  & 0.07 & 0.00 & 0.12 & 0.18 \\
LangGraph-Star & Strict & 0.04 & 0.00 & 0.07 & 0.04 \\
\midrule
AutoGen-Mesh   & None   & 0.26 & 0.10 & 0.68 & 0.76 \\
AutoGen-Mesh   & Speed  & 0.05 & 0.00 & 0.15 & 0.17 \\
AutoGen-Mesh   & Strict & 0.05 & 0.00 & 0.15 & 0.08 \\
\bottomrule
\end{tabular}
\end{table}

As shown in Table~\ref{tab:compact_threshold_diagnostics}, both settings exhibit a clear reduction in fitted transmission strength after governance is enabled. In LangGraph-Star, Speed reduces $\beta$ from 0.42 to 0.07 and lowers the observed final coverage from 0.80 to 0.18, while Strict further reduces the final coverage to 0.04. In AutoGen-Mesh, Speed and Strict reduce $\beta$ from 0.26 to 0.05, with the observed final coverage decreasing from 0.76 to 0.17 and 0.08, respectively. These results provide additional fitted evidence for the trajectory-level results: the governance layer reduces effective propagation strength and keeps finite-horizon infection coverage low under representative defense settings.
\section{SUPPLEMENT TO VULNERABILITY}
\label{app:threat-model-impl}

\vspace{3pt}

\noindent \textbf{Communication Structures} We model each framework as a discrete-time process that repeatedly selects an acting agent, exposes a context view to that agent, and appends the produced message or artifact to a shared state that may be visible to others in later steps. The induced visibility pattern defines the effective communication structure and the path along which an injected seed can propagate.

\noindent \textbf{For LangChain}, we use a sequential, single-pass pipeline in which roles execute in a fixed order and each step consumes the accumulated upstream outputs as context. Information flows strictly forward, with no backward dependency from later roles to earlier roles, yielding a directed chain. In our tasks this corresponds to a four-stage pipeline with roles such as product manager, architect, engineer, and reviewer, where the reviewer only sees artifacts that have been produced earlier in the chain. This mapping matches the intended sequential composition model, where intermediate results are passed downstream as the next component's input.

\noindent \textbf{For MetaGPT}, we use an SOP-style workflow in which roles advance through staged production of intermediate artifacts, and downstream roles read multiple upstream artifacts as part of their context view. Execution remains stage-ordered, but visibility is cumulative across artifacts rather than restricted to a single immediate predecessor, yielding a chain-like structure with cross-stage carryover. In our setup, SOP stages align with roles such as product manager, architect, engineer, and QA/reviewer, and the reviewer consumes the PRD, design, and code artifacts produced in earlier stages. This mapping matches MetaGPT's core design, where SOP stages and their artifacts are first-class objects reused in subsequent stages.

\noindent \textbf{For CrewAI}, we use a hierarchical process in which a manager agent issues directives, workers respond under those directives, and the manager aggregates worker outputs before issuing the next directive. Workers primarily interact through the manager rather than directly with one another, producing a hub-and-spoke structure over rounds. In our configuration, the crew contains a manager and worker roles such as researcher, developer, and reviewer, with the reviewer treated as a worker who responds to manager instructions but does not directly read other workers' raw messages. This mapping matches the hierarchical delegation semantics, where a manager coordinates and validates worker contributions.

\noindent \textbf{For LangGraph}, we use a supervisor-worker team where a supervisor node routes control to a selected worker, receives the worker's result, updates the shared state, and continues routing. The supervisor maintains the global view of the evolving state, while only the activated worker executes the next step, yielding a centralized routing structure with alternating supervisor-to-worker transitions. In our tasks, workers include roles such as planner, coder, and reviewer, and the reviewer is activated when the supervisor routes evaluation or checking work to that node based on the current state. This mapping matches the supervisor routing pattern used to construct hierarchical agent teams.

\begin{table}[t]
\caption{Cascade severity by topology.}
\vspace{2pt}
\label{tab:cascade_stats}
\centering
\resizebox{0.95\columnwidth}{!}{
    \begin{tabular}{llccc}
    \toprule
    \textbf{Framework} & \textbf{Topology} & \textbf{Agents} & \textbf{Final Rate} & \textbf{Std} \\
    \midrule
    MetaGPT      & Chain & 4 & 100.0\% & 0.0\% \\
    LangChain    & Chain & 4 & 89.2\%           & 22.1\% \\
    LangGraph    & Star  & 4 & 100.0\% & 0.0\% \\
    CrewAI       & Star  & 4 & 100.0\% & 0.0\% \\
    AutoGen      & Mesh  & 3 & 100.0\% & 0.0\% \\
    CAMEL        & Mesh  & 3 & 100.0\% & 0.0\% \\
    \bottomrule
    \end{tabular}
}
\end{table}

\noindent \textbf{For AutoGen}, we use a group chat in which agents share a broadcast conversation history, while the next speaker is selected dynamically from the same shared history. Each produced message is appended to the shared state and becomes visible to all participants, yielding a mesh under broadcast visibility with a data-dependent activation schedule. In our configuration, the group contains agents such as a task coordinator, coder, and reviewer, and the reviewer reads the full conversation history before issuing comments or approval. This mapping matches the group-chat semantics that support model-based next-speaker selection under shared context.

\noindent \textbf{For CAMEL}, we use a role-playing, multi-round dialogue collaboration in which role agents iteratively contribute messages to a shared dialogue state. The essential property is that the evolving dialogue history is the primary context for subsequent turns, yielding a mesh-like interaction under shared history; the concrete speaking order is treated as an operational choice that does not change the role-playing dialogue semantics. In our experiments, roles include a task giver, an assistant coder, and a reviewer, and the reviewer participates as a dialogue agent that inspects the shared history and comments on intermediate solutions. This mapping matches CAMEL's role-playing paradigm, where collaboration proceeds through iterative dialogue among role agents.

\vspace{3pt}

\noindent \textbf{Injection Definition and Node Mapping} For the endogenous propagation study, we realize the adversary as a single injected seed applied to a designated target agent by prefixing a fixed instruction to that agent's system-level directive at the first step. This setting is used only to standardize the initial error condition for measuring endogenous amplification, and it is separated from the later application-layer attack instantiation. Injection locations are normalized as \emph{entry}, \emph{hub}, and \emph{leaf}: entry denotes the earliest upstream role in chain or SOP workflows, hub denotes the central coordinator in star workflows, and leaf denotes a non-coordinator worker role such as an engineer or reviewer.

\vspace{3pt}

\noindent \textbf{Observables and Counting Rules} We log a binary infection state $X_{i,r}(t)$ for agent $i$ in run $r$ at step $t$ by tracking the injected canonical seed in controlled traces. A state is counted as adopted only when the seed is carried forward as part of the agent output, shared context, or downstream artifact; rejected or explicitly negated mentions are excluded. Beyond final outcomes, we record polluted intermediate artifacts (PIA): an intermediate message or artifact is flagged as PIA if it is committed to the shared state, reused in downstream context, and classified as adopted-infected by the same controlled-seed counting rule.

\vspace{3pt}

\noindent \textbf{Extended Severity Statistics for Cascade Amplification} Table~\ref{tab:cascade_stats} reports the final infection rate and its standard deviation for the cascade amplification experiment in each framework.

\begin{table}[!t]
\centering
\normalsize
\caption{Framework instantiations used in Section~\ref{sec:eval-setup}.}
\label{tab:framework_budget}
\vspace{3pt}
\setlength{\tabcolsep}{6pt}
\renewcommand{\arraystretch}{1.08}

\begin{tabular*}{0.75\columnwidth}{@{\extracolsep{\fill}}llc@{}}
\toprule
\textbf{Framework} & \textbf{Topology} & \textbf{Agents} \\
\midrule
LangChain & chain & 4 \\
MetaGPT   & chain & 4 \\
AutoGen   & mesh  & 4 \\
CAMEL     & mesh  & 4 \\
CrewAI    & star  & 4 \\
LangGraph & star  & 4 \\
\bottomrule
\end{tabular*}
\end{table}

\section{Supplement to Evaluation}
\label{app:appendix-c}

\vspace{3pt}
\noindent \textbf{Tasks and application-level adaptation.}
We evaluate three scenarios: \textsc{Quant} (constructed from the UCI repository~\cite{uci}), \textsc{Rigid} (derived from MATH-style multi-step reasoning problems~\cite{math}), and \textsc{MMLU} (adapted from MMLU~\cite{mmlu} into retrieval-based question answering). 

We apply an application-level adaptation that requires agents to invoke external tools or retrieve external resources during problem solving, which operationalizes a supply-chain facing attack surface. \textsc{Quant} is adapted from the UCI Adult Income dataset: each instance explicitly requires loading a remote CSV and completing the specified analysis through a Python data-analysis toolchain rather than answering from memory. \textsc{Rigid} is adapted from AMC 10/12-style number theory and probability problems: agents are required to call \texttt{MathAPI} (or a designated calculator tool) to validate each step before proceeding. \textsc{MMLU} uses a subset from High School Physics, Chemistry, and Biology, and is converted into a retrieval-based setting backed by a \texttt{Wikipedia API}; each question is paired with a single target fact, and the task requires retrieving and verifying that fact from the external knowledge source prior to producing the final answer.

Each scenario contains 20 task instances. In the attack evaluation, each framework--scenario--attacker-policy cell is repeated three times, yielding $20 \times 3$ runs per cell. In the defense evaluation, each selected framework--scenario--defense-method--attacker-policy cell is repeated twice, yielding $20 \times 2$ runs per cell. 

\begin{table}[htbp]
\centering
\normalsize
\caption{NLI comparison results on the REAL evaluation set.}
\vspace{3pt}
\label{tab:nli_comparison}
\setlength{\tabcolsep}{3pt}
\renewcommand{\arraystretch}{1.05}

\begin{tabular}{lccccc}
\toprule
\textbf{Method} & \textbf{Acc.} & \textbf{FGR} & \textbf{G-Prec.} & \textbf{Time} & \textbf{Tok.} \\
\midrule
gpt-4o-mini          & 78.8\% & 20.5\% & 43.8\%  & 684.4ms & 3993 \\
deberta-v3-small     & 71.2\% & 2.3\%  & 80.0\%  & 148.2ms & 0 \\
deberta-v3-large     & 82.7\% & 0.0\%  & 100.0\% & 91.1ms  & 0 \\
similarity (Jaccard) & 11.5\% & 0.0\%  & N/A     & 0.0ms   & 0 \\
\bottomrule
\end{tabular}

\end{table}

\vspace{2pt}
\noindent \textbf{Baseline Setup.}
We adapt three comparison baselines under the same task, framework, topology, prompt, attack injection, and evaluation detector settings. Reflection serves as a sender-side self-check baseline: before message delivery, the agent reviews and revises its own output. AGrail is used as an agent-monitoring baseline placed on the inter-agent message path~\cite{luo2025agrail}. It checks each outgoing message using task context, role information, safety criteria, and safety-check memory; messages judged unsafe are replaced with a guardrail placeholder before entering downstream context. The CFG baseline follows the graph-control setting inspired by ControlValve~\cite{jha2025breaking}. It derives admissible routing rules from the agent list, topology, allowed tools, and task instruction, then checks runtime agent transitions against these rules and supports bounded rejection or re-planning. These baselines receive no extra oracle information beyond the normal task text and data source.

\vspace{2pt}
\noindent \textbf{Prompt.} The detailed prompt templates can be found in our open-source code repository.

\vspace{2pt}
\noindent \textbf{Framework instantiation budget.}
Appendix~\ref{app:threat-model-impl} specifies the agent rosters, role names, message routing, and termination criteria for each framework instantiation. Here we only summarize the agent count used in our experiments.

\vspace{2pt}
\noindent \textbf{Attacker policies and injection protocol.}
We instantiate three application-level attacker policies via a single message injection. \textsc{Baseline} inserts the malicious claim verbatim; \textsc{Compliance} frames the claim as a cooperative instruction aligned with the local task objective; \textsc{Security\_FUD} frames the claim as a security-motivated caution that biases downstream decisions. For each run, we inject exactly one seed message at a deterministic injection target determined by the orchestration in Appendix~\ref{app:threat-model-impl}, then measure whether multi-round collaboration propagates the seed into an infected final artifact.

\vspace{3pt}
\noindent \textbf{NLI module evaluation for conservative Green screening.}
The local lightweight NLI model is not intended to be the strongest three-way classifier. In the defense module, it is used as a conservative first-pass screen for Green labels, where the main risk is allowing unsupported claims to pass as entailment. We evaluate this behavior on a human-annotated set whose core samples are extracted from real collaboration logs of \textsc{MetaGPT}, \textsc{LangGraph}, and \textsc{AutoGen}, covering intermediate factual statements produced under \textsc{Quant} and \textsc{Rigid}. Because real traces are typically dominated by \texttt{neutral}, we add a small amount of synthetic augmentation by writing entailment paraphrases and contradiction variants (numerical conflicts) for selected ground-truth facts, then manually checking all labels against the underlying evidence. FGR is the share of non-entailment samples predicted as entailment (Green). Green Precision is the share of predicted Green samples that are truly entailment, and is undefined when no Green prediction is made. We use DeBERTa-v3-small as the default online NLI backend under local compute constraints.

\begin{table*}[t]
\centering
\small
\setlength{\tabcolsep}{7pt}
\renewcommand{\arraystretch}{1.15}
\caption{Per-dataset aggregate results across defense methods.}
\label{tab:dataset_aggregate_main}

\begin{tabular*}{0.92\textwidth}{@{\extracolsep{\fill}}l ccc ccc ccc@{}}
\toprule
\multirow{2}{*}{\textbf{Defense}} &
\multicolumn{3}{c}{\textbf{BICR (\%) $\uparrow$}} &
\multicolumn{3}{c}{\textbf{Safe Completion (\%) $\uparrow$}} &
\multicolumn{3}{c}{\textbf{Tok./Safe ($10^3$) $\downarrow$}} \\
\cmidrule(lr){2-4}
\cmidrule(lr){5-7}
\cmidrule(lr){8-10}
& \textbf{M} & \textbf{Q} & \textbf{R}
& \textbf{M} & \textbf{Q} & \textbf{R}
& \textbf{M} & \textbf{Q} & \textbf{R} \\
\midrule
Strict     & 89.4 & 98.9 & 94.4 & 88.3 & 98.9 & 90.6 & 54.9 & 63.2 & 50.9 \\
Balanced   & 87.2 & 97.5 & 93.9 & 84.2 & 97.2 & 92.2 & 29.2 & 36.1 & 27.0 \\
Speed      & 92.8 & 83.1 & 92.5 & 91.1 & 83.1 & 89.7 & 20.1 & 23.4 & 18.7 \\
Reflection & 35.3 & 29.7 & 31.9 & 35.0 & 29.7 & 30.3 & 12.8 & 12.9 & 12.4 \\
AGrail     & 82.5 & 75.3 & 78.6 & 20.8 & 7.2  & 5.3  & 23.5 & 21.8 & 21.2 \\
CFG        & 70.6 & 79.4 & 78.9 & 25.8 & 15.8 & 7.2  & 17.0 & 18.4 & 17.1 \\
\bottomrule
\end{tabular*}

\end{table*}

\section{Per-Dataset Aggregate Defense Results}
\label{app:full_dataset_results}

Table~\ref{tab:dataset_aggregate_main} reports the aggregate defense results on each evaluation dataset. Values are averaged over frameworks and attack settings within each dataset.

\end{document}